\documentclass[12pt]{article}

\usepackage{amsmath}
\usepackage{amssymb}
\usepackage{xspace}
\usepackage{gensymb}
\usepackage{graphicx}
\usepackage{setspace}
\usepackage{hyperref}
\usepackage{multirow}
\usepackage{xcolor}
\usepackage{verbatim}
\usepackage{upgreek}
\usepackage{booktabs}

\usepackage{scicite}

\usepackage{times}

\newenvironment{sciabstract}{%
\begin{quote} \bf}
{\end{quote}}

\newcounter{lastnote}

\def\titlestring{A magnified compact galaxy at redshift 9.51 with strong nebular emission lines}

\title{\titlestring} 

\author{Hayley Williams$^{1}$, Patrick L. Kelly$^{1}$, Wenlei Chen$^{1}$, Gabriel Brammer$^{2}$,\\ 
Adi Zitrin$^{3}$, Tommaso Treu$^{4}$, Claudia Scarlata$^{1}$, \\
Anton M. Koekemoer$^{5}$, Masamune Oguri$^{6,7}$, Yu-Heng Lin$^{1}$,\\
Jose M. Diego$^{8}$, Mario Nonino$^{9}$, Jens Hjorth$^{10}$, Danial Langeroodi$^{10}$, \\
Tom Broadhurst$^{11}$, Noah Rogers$^{1}$, Ismael Perez-Fournon$^{12,13}$,\\
Ryan J. Foley$^{14}$, Saurabh Jha$^{15}$, Alexei V. Filippenko$^{16}$,\\
Lou Strolger$^{5}$, Justin Pierel$^{5}$, Frederick Poidevin$^{12,13}$, and Lilan Yang$^{17}$\\
\footnotesize{$^{1}$Minnesota Institute for Astrophysics, University of Minnesota,  Minneapolis, MN 55455, USA}\\
\footnotesize{$^{2}$Cosmic Dawn Center, Niels Bohr Institute, University of Copenhagen, DK-2200 Copenhagen, Denmark}\\
\footnotesize{$^{3}$Physics Department, Ben-Gurion University of the Negev, Beer-Sheva 8410501, Israel}\\
\footnotesize{$^{4}$Department of Physics and Astronomy, University of California, Los Angeles, CA 90095, USA}\\
\footnotesize{$^{5}$Space Telescope Science Institute, Baltimore, MD 21218, USA}\\
\footnotesize{$^{6}$Center for Frontier Science, Chiba University, Chiba 263-8522, Japan}\\
\footnotesize{$^{7}$Department of Physics, Chiba University, Chiba 263-8522, Japan}\\ 
\footnotesize{$^{8}$Instituto de F\'isica de Cantabria, Universidad de Cantabria, Consejo Superior de }\\
\footnotesize{ Investigaciones Cient\'ificas, 39005 Santander, Spain}\\
\footnotesize{$^{9}$Istituto Nazionale di Astrofisica, Osservatorio Astronomico di Trieste, 34124 Trieste, Italy}\\
\footnotesize{$^{10}$Dark Cosmology Center, Niels Bohr Institute, University of Copenhagen, DK-2200 Copenhagen, Denmark }\\
\footnotesize{$^{11}$Donostia International Physics Center, Ikerbasque Foundation, }\\
\footnotesize{University of the Basque Country, 20018 Donostia, Spain}\\
\footnotesize{$^{12}$Instituto de Astrofisica de Canarias, E-38205 La Laguna, Tenerife, Spain}\\
\footnotesize{$^{13}$Departamento de Astrof\'{\i}sica, Universidad de La Laguna, 38206 La Laguna, Tenerife, Spain}\\
\footnotesize{$^{14}$Department of Astronomy and Astrophysics, University of California Observatories/Lick Observatory, }\\
\footnotesize{University of California, Santa Cruz, CA 95064, USA}\\
\footnotesize{$^{15}$Department of Physics and Astronomy, Rutgers, The State University of New Jersey, Piscataway,}\\
\footnotesize{NJ 08854, USA}
}

\date{}

\begin{document} 


\baselineskip24pt


\maketitle

\begin{centering}
\linespread{1.37}
\footnotesize{$^{16}$Department of Astronomy, University of California, Berkeley, CA 94720-3411, USA}\\
\footnotesize{$^{17}$Kavli Institute for the Physics and Mathematics of the Universe, The University of Tokyo, Kashiwa, Japan 277-8583}\\
\linespread{1}
\end{centering}






\begin{sciabstract}
Ultraviolet light from early galaxies is thought to have ionized the intergalactic gas. However, there are few observational constraints on this epoch, due to the faintness of those galaxies and the redshifting of their light in to the infrared. We report the discovery, in JWST imaging, of a distant galaxy that is magnified by gravitational lensing. JWST spectroscopy, at rest-frame optical wavelengths, detects strong nebular emission lines that are attributable to oxygen and hydrogen. The measured redshift is $z= 9.51 \pm 0.01$, corresponding to 510 million years after the Big Bang. The galaxy has a radius of 16.2$^{+4.6}_{-7.2}$ parsecs, which is substantially more compact than galaxies with equivalent luminosity at $z\sim$ 6 to 8, leading to a high star formation rate surface density.

\end{sciabstract}




Radiation from early galaxies is thought to be responsible for the reionization of the Universe, the process in which the majority of the intergalactic neutral gas was ionized by high energy photons. Observational constraints suggest that reionization was completed when the Universe was approximately one billion years old (redshift $z\sim6$) \cite{Planck_2018}. The precise timeline of reionization, and the relative contributions of faint and bright galaxies to the ionizing photon budget, remain uncertain \cite{Finkelstein_2019}. Observations of distant galaxies that existed during the epoch of reionization are crucial to understanding the physical processes that occured during that period \cite{Dayal_2019}. 

The intrinsic faintness and small angular sizes of galaxies at high redshifts constrain our ability to observe them in detail. Because of their very large masses, galaxy clusters act as gravitational lenses, magnifying the flux and stretching the angular extent of distant background galaxies. Gravitational lensing can therefore extend the observational limits of a telescope, probing faint and small galaxies at high redshifts that would otherwise be undetectable \cite{Robertson_2013}. 

Near-infrared imaging has identified distant galaxy candidates at $z\gtrsim9$ and up to $z\simeq17$ \cite{Adams_2022,  Castellano_2022,Yan_2022}, but the redshifts of those candidates have not been confirmed with spectroscopy. Among these candidates are an unexpectedly large number of galaxies with bright ultraviolet (UV) absolute magnitudes ($M_{\rm UV}\lesssim -21$~mag; \cite{Atek_2022,Finkelstein_2022, Naidu_2022}) and high stellar masses ($M_{\ast} > 10^{10}$ solar masses (${\rm M}_{\odot}$); \cite{Labbe_2022}). This population was not predicted by simulations of early galaxy formation that assumed standard cosmology \cite{Mason_2022,Boylan-Kolchin_2022}. Spectroscopy is necessary to confirm the redshifts of these galaxies and infer their physical properties from the strengths of their emission lines. 

Nebular emission lines are produced by clouds of interstellar gas within a galaxy, and spectroscopic analysis of these lines can provide information about the density, temperature, and chemical composition of the gas. Spectroscopy has confirmed three high-redshift galaxies ($7.66 < z < 8.50$) with detections of strong nebular emission lines \cite{Carnall_2022} and the temperature-sensitive  [O~\textsc{iii}] 4363 \AA $ $ emission line, which has been used to make the first direct electron temperature oxygen abundance measurements in galaxies at these redshifts \cite{Arellano_Cordova_2022, Curti_2022, Rhoads_2023,Schaerer_2022,Trump_2022}. There has been further spectroscopic confirmation of seven galaxies from $z=7.762$ – $8.998$ \cite{Fujimoto_2023}.

\begin{figure*}
     \centering
     \includegraphics[width=5.0in]{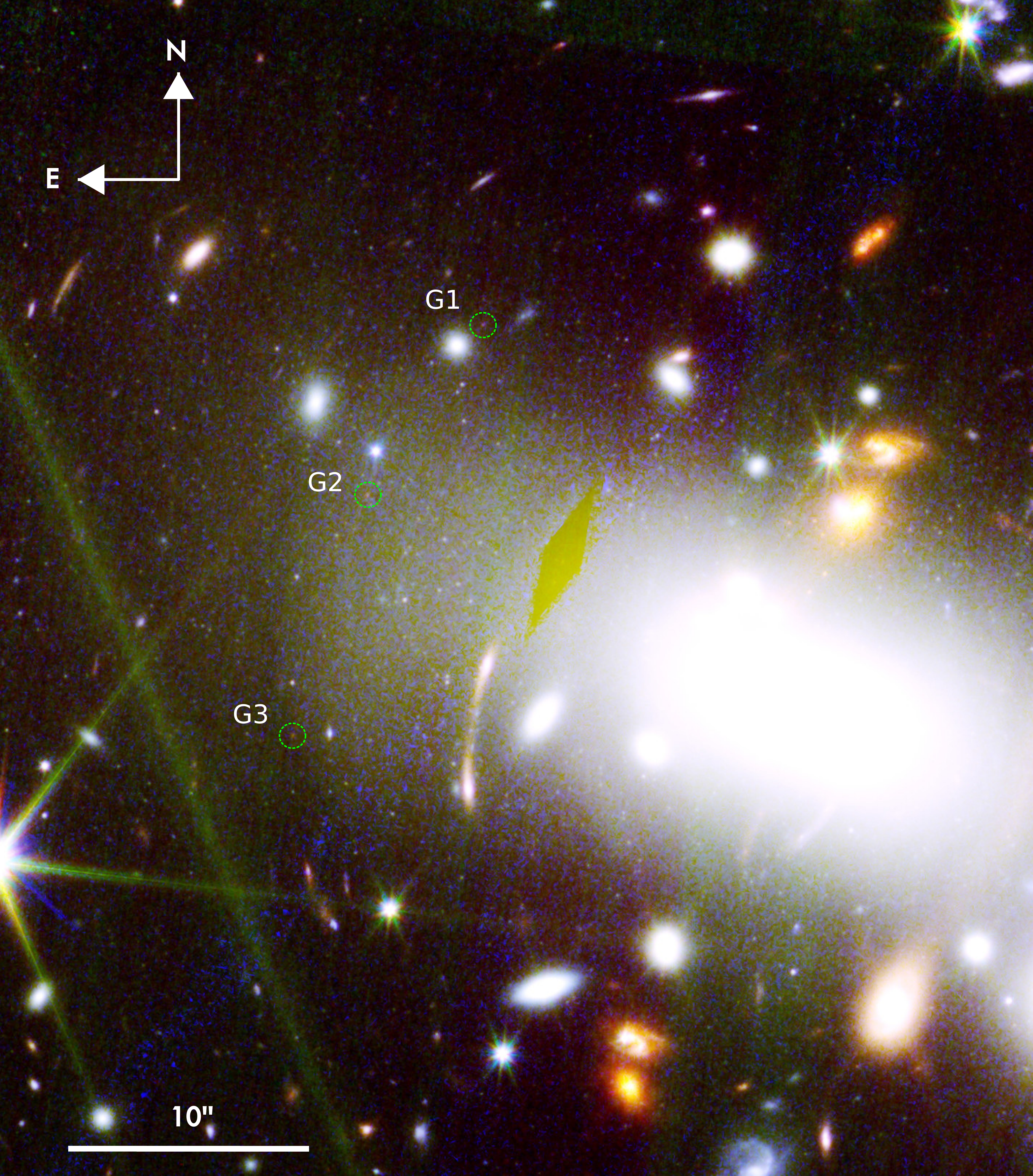}
     \label{fig:11027_img}
    \caption{\textbf{Color-Composite Image of Part of RX\,J2129 }JWST + HST color-composite image of galaxy cluster RX\,J2129, with three images of the $z=9.51$ galaxy circled in green. We observed image G2 with NIRSpec. Red: JWST F277W+F356W+F444W, Green: JWST F115W+F150W+F200W, Blue: HST ACS F606W + F814W.  The broad blue and green bands are diffraction spikes from foreground stars. The yellow diamond is an artefact caused by a chip gap in the HST ACS camera. The individual red, green, and blue images are shown in Figures \ref{fig:rxj2129_red}, \ref{fig:rxj2129_green}, and \ref{fig:rxj2129_blue}.}
    \label{fig:imglarge}
\end{figure*}

\paragraph*{Imaging observations and analysis}\label{sec:data}
We obtained imaging of the RX\,J2129+0005 cluster field (hereafter RX\,J2129) on 2022 October 6 with the Near-Infrared Camera (NIRCam) instrument on JWST operating in imaging mode as part of a Director's Discretionary program (DD 2767; PI P. Kelly). NIRCam spans a wavelength range of 0.6~$\upmu$m to 5.0~$\upmu$m; we obtained exposures in the F115W, F150W, F200W, F277W, F356W, and F444W filters. The name of each filter indicates its approximate central wavelength and bandwidth; for example, the central wavelength of the F115W filter is approximately 1.15~$\upmu$m and the bandwidth is wide (0.225~$\upmu$m). Our exposure times ranged from 2026~s for the F444W filter to 19927~s for the F150W filter. The astrometric alignment for the NIRCam images was performed using a catalog prepared from previous imaging taken with the Suprime-Cam instrument on the Subaru telescope \cite{Supplementals}. The color-composite NIRCam image of the RX\,J2129 cluster is shown in Figure \ref{fig:imglarge}. In this image, we identified a candidate distant galaxy (designated RX\,J2129-z95) which appears as three images (designated RX\,J2129-z95:G1, RX\,J2129-z95:G2, and RX\,J2129-z95:G3) due to the gravitational lensing of the foreground cluster. Coordinates for the three images of RX\,J2129-z95 are given in Table \ref{tab:coordinates}. Photometric measurements from the NIRCam imaging, along with measurements from previous Hubble Space Telescope (HST) imaging of the RX\,J2129 cluster field obtained with the Advanced Camera for Surveys (ACS) and the Wide Field Camera 3 (WFC3), are listed in Table \ref{tab:phot_measurements} \cite{Supplementals}.  

\begin{figure*}
     \centering

    \includegraphics[width=1\linewidth]{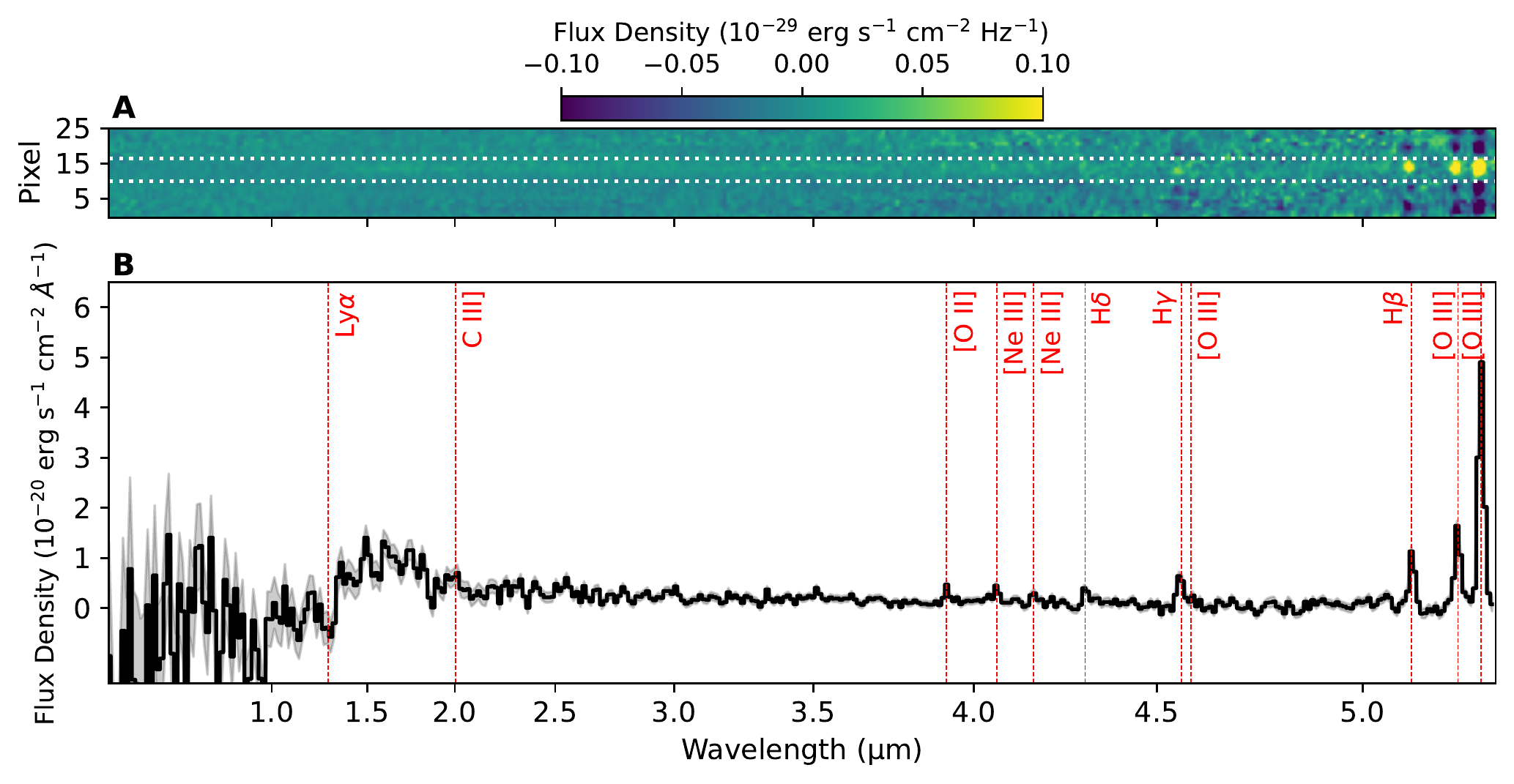}

    \caption{\textbf{Observed JWST spectrum of image G2 }NIRSpec prism (A) 2D and (B) 1D spectra of image G2 of the $z=9.51$ galaxy. In panel B, the black line is the data, and the red vertical lines indicate the expected observed wavelengths of emission lines at $z=9.51$. Grey shading indicates 1$\sigma$ uncertainties. In panel A, the apparent negative fluxes in the background near the emission lines in the 2D spectrum results from the dither pattern used for the NIRSpec observations. The white dotted lines show the extraction window for the galaxy. This spectrum has not been corrected for magnification due to gravitational lensing.}
    \label{fig:full_spec}
    
\end{figure*}

We used the \textsc{eazy-py} software \cite{Brammer_2008} to constrain the photometric redshift (an estimate for a source's redshift made without the use of spectroscopy) of all sources in the field detected in the NIRCam imaging \cite{Supplementals}. We obtained a photometric redshift of $z_\textrm{phot}=9.38^{+0.29}_{-0.15}$ for image G2 of RX\,J2129-z95. From the NIRCam photometry, we estimate a UV spectral slope ($\beta$) of -1.98 $\pm$ 0.11 \cite{Supplementals}. Using the F150W photometric flux measurement and correcting for the effect of magnification due to gravitational lensing on image G2 (magnification $\mu=20.2\pm3.8$ \cite{Supplementals}), we calculate the absolute UV magnitude at 1500~\AA \ $M_\textrm{UV} = -17.4 \pm 0.22$ mag. 

We use the \textsc{prospector} software \cite{Johnson_2021} to infer the physical properties of the galaxy from the spectral energy distribution (SED) of image G2, using the NIRCam photometry and non-detections from archival optical HST imaging \cite{Supplementals}. Before doing so, we correct the photometry for the effect of magnification due to gravitational lensing. We find that the galaxy has a low stellar mass $\log(M_*/{\rm M}_{\odot})=7.63^{+0.22}_{-0.24}$ (uncertainty is 1$\sigma$ and includes the propagated uncertainty in magnification). The template fitting also indicates an oxygen abundance of $12+\log(\textrm{O}/\textrm{H}) = 7.63^{+.07}_{-.05}$. The best-fitting star formation history (SFH) has a mass-weighted age of $56^{+43}_{-34}$~Myrs and indicates a star formation rate (SFR) of SFR$ = 0.9 \pm 0.32$ M$_\odot$ yr$^{-1}$. The observed SED of image G2 and best fitting \textsc{prospector} model are shown in Figure \ref{fig:photometry_11027} \cite{Supplementals}.

We use the \textsc{Lenstruction} software \cite{Yang2020, Birrer2021} to reconstruct the F150W image G2 of the galaxy, correcting for the effects of gravitational lensing and convolution with the NIRCam point spread function (PSF). We fit the reconstructed image with a surface brightness model, consisting of an elliptical S\`{e}rsic profile with index $n$ fixed to 0.5. $n$ determines the degree of curvature of the profile, and $n=0.5$ represents a Gaussian profile. This indicates the intrinsic half-light radius of the reconstructed source $R_{\textrm{e, intrinsic}}=16.2^{+4.6}_{-7.2}$ pc (Figure \ref{fig:imfit}). We also fit the observed F150W image directly, using the \textsc{Galight} software \cite{Ding2020}, which indicates an observed angular size of $\theta_\textrm{e, observed}=0.04\pm0.01$ arcseconds. 

\paragraph*{Spectroscopic Observations and Analysis}\label{sec:data} We obtained follow-up spectroscopy of the RX\,J2129 cluster field on 2022 October 22, using JWST's Near Infrared Spectrograph (NIRSpec) in Multi-Object Spectroscopy (MOS) mode. Targets were selected based on photometric redshift estimates from the NIRCam imaging. We used a standard 3-shutter dither pattern and obtained a 4464~s exposure using the prism disperser. This setup provides wavelength coverage from 0.6~$\upmu$m to 5.3~$\upmu$m, with spectral resolving power $R$ ranging from $\approx50$ to $\approx400$ \cite{Jakobsen_2022}. The fully calibrated one-dimensional (1D) and two-dimensional (2D) spectra are shown in Figure \ref{fig:full_spec} \cite{Supplementals}.  

We estimate the spectroscopic redshift of the galaxy by visual identification of the emission lines H$\beta$ and [O~\textsc{iii}] 4959,5007 \AA. We refine our redshift measurement by modeling the emission-line profiles (see below), which yields $z_{\textrm{spec}}=9.51 \pm 0.01$. The spectroscopic redshift is consistent with the photometric redshift ($z_{\textrm{phot}}=9.38^{+0.29}_{-0.15}$), indicating that the lines have not been misidentified. 

To constrain the fluxes of the emission lines, we use the \textsc{pPXF} (Penalized Pixel-Fitting) software \cite{Cappellari_2022}, which models the stellar continuum and fits Gaussian profiles to each of the emission lines \cite{Supplementals}. Table \ref{tab:fluxes} lists our measured emission-line fluxes, equivalent widths (EWs), and corresponding uncertainties. We do not detect Ly$\alpha$, with a 3$\sigma$ upper limit for its flux of approximately $39\times10^{-19}$ erg s$^{-1}$ cm$^{-2}$ \cite{Supplementals}.  We assume negligible extinction due to dust and apply no reddening correction to the flux measurements \cite{Supplementals}.

\begin{table}
\centering
\setlength{\tabcolsep}{14pt}
\caption{\textbf{Emission Line Flux Measurements. } Flux measurements and rest frame EWs of emission lines for the $z=9.51$ galaxy. The flux measurements have not been corrected for magnification due to gravitational lensing. Upper limits are 3$\sigma$.}
\begin{tabular}{@{}lccc@{}}
\\

& \textbf{Rest Frame} & \textbf{Observed Flux} & \textbf{Rest Frame} \\
 \textbf{Emission Line} & \textbf{Wavelength (\AA)} & \textbf{(10$^{-19}$ erg s$^{-1}$ cm$^{-2}$)}&\textbf{EW (\AA)}\\
\midrule
Ly$\alpha$ & 1216 \ & \textless~39 & \textless~31\\
C~\textsc{iii}] + [C~\textsc{iii}] &1907, 1909 \ & \textless~20 & \textless~51\\
{[O~\textsc{ii}]} & 3626, 3629& 5.9 $\pm$ 1.6 & 44.0 $\pm$ 11.7\\
{[Ne~\textsc{iii}]} & 3869 &6.3 $\pm$ 1.4 & 53.4 $\pm$ 11.6 \\
{[Ne~\textsc{iii}]} + H$\epsilon$ & 3968, 3970 & \textless~4.9 & \textless~39\\
H$\delta$ & 4102 & 5.7 $\pm$ 1.2 & 52.1 $\pm$ 11.0\\
H$\gamma$ & 4340 & 11.8 $\pm$ 1.7 & 194.9 $\pm$ 27.0\\
{[O~\textsc{iii}]} & 4363 & \textless~5.0 & \textless~74 \\
H$\beta$ & 4861 & 17.8 $\pm$ 2.5 & 248.0 $\pm$ 35.0\\
{[O~\textsc{iii}]} & 4959 & 26.3 $\pm$ 1.8 & 392.3 $\pm$ 26.8 \\
{[O~\textsc{iii}]} & 5007 & 79.0$\pm$ 2.0 & 1092.0$\pm$ 28.4\\
\bottomrule

\end{tabular}

\label{tab:fluxes}
\end{table}

We infer the star-formation rate (SFR) of the galaxy from our H$\beta$ flux measurement using the relation,
\begin{equation}
    \textrm{SFR} \; / \; ({\rm M}_\odot \; \textrm{yr}^{-1}) = 5.5 \times 10^{-42} \: L({\rm H}\alpha) \; / \; (\textrm{erg} \; \textrm{s}^{-1})\, ,
\end{equation}
\noindent
where $L({\rm H}\alpha)$  is the intrinsic H$\alpha$ luminosity of the galaxy. To compute $L({\rm H}\alpha)$, we correct for magnification due to lensing and assume Case B recombination \cite{Osterbrock_1989}. We find SFR = $1.69^{+0.51}_{-0.34} $~M$_\odot$~yr$^{-1}$ \cite{Supplementals}. This value is approximately 50\% larger than the value we derived from the SED fitting ($0.90 \pm 0.32$ M$_\odot$~yr$^{-1}$), but the discrepancy is $<2\sigma$. Using the stellar mass we inferred from the SED fitting ($\log(M_*/{\rm M}_{\odot})=7.63^{+0.22}_{-0.24}$), we compute the specific star formation rate (sSFR, the SFR per unit mass). We find $\log(\textrm{sSFR})=-7.38\pm0.26$ yr$^{-1}$. 

To test for a spatial offset between the nebular emission and the stellar continuum, we extract profiles along the spatial axis of the NIRSpec MOS slit. We extract spatial profiles of the strong emission lines in 0.05~$\upmu$m windows. For the stellar continuum, we extract the spatial profile of the spectrum at all wavelengths above 1.5 $\mu$m, masking out the regions within 0.05~$\upmu$m of any strong emission lines. We find no evidence for an offset between the nebular emission lines and stellar continuum (Figure \ref{fig:profiles}). 

We use the fluxes of the strongest emission lines of oxygen and hydrogen to estimate the oxygen abundance of the $z=9.51$ galaxy. The high ratio $\textrm{O}_{32} \equiv F([\textrm{O}~\textsc{iii}])/F([\textrm{O}~\textsc{ii}]) =13 \pm 4$ we calculate for this galaxy is consistent with highly ionized gas with low metallicity. We therefore use an empirical calibration derived by \cite{Izotov2019} measured from low-metallicity ($12 + \log(\textrm{O}/\textrm{H}) \lesssim 8.0$) galaxies,
\begin{equation}
12 + \log(\textrm{O}/\textrm{H}) = 0.950 \log(\textrm{R}_{23}-0.08\textrm{O}_{32}) +6.805
\end{equation}
where $\textrm{R}_{23} \equiv (F([\textrm{O}~\textsc{ii}] 3727~\textrm{\AA}) \ + F([\textrm{O}~\textsc{iii}] 4959~\textrm{\AA}) \ + F([\textrm{O}~\textsc{iii}] 5007~\textrm{\AA} )) / F(\textrm{H}\beta)$. For the $z=9.51$ galaxy, we find an oxygen abundance of 12 + log(O/H) $=7.48\pm 0.08$, where the uncertainty includes both line-flux and calibration uncertainties \cite{Supplementals}. Using other calibrations \cite{Maiolino2008, Jiang_2019} results in consistent estimates. The oxygen abundance derived from the photometry is also consistent with the emission line calibrations within 1.5$\sigma$ \cite{Supplementals}.

\begin{figure}
    \centering
    \includegraphics[width=\textwidth]{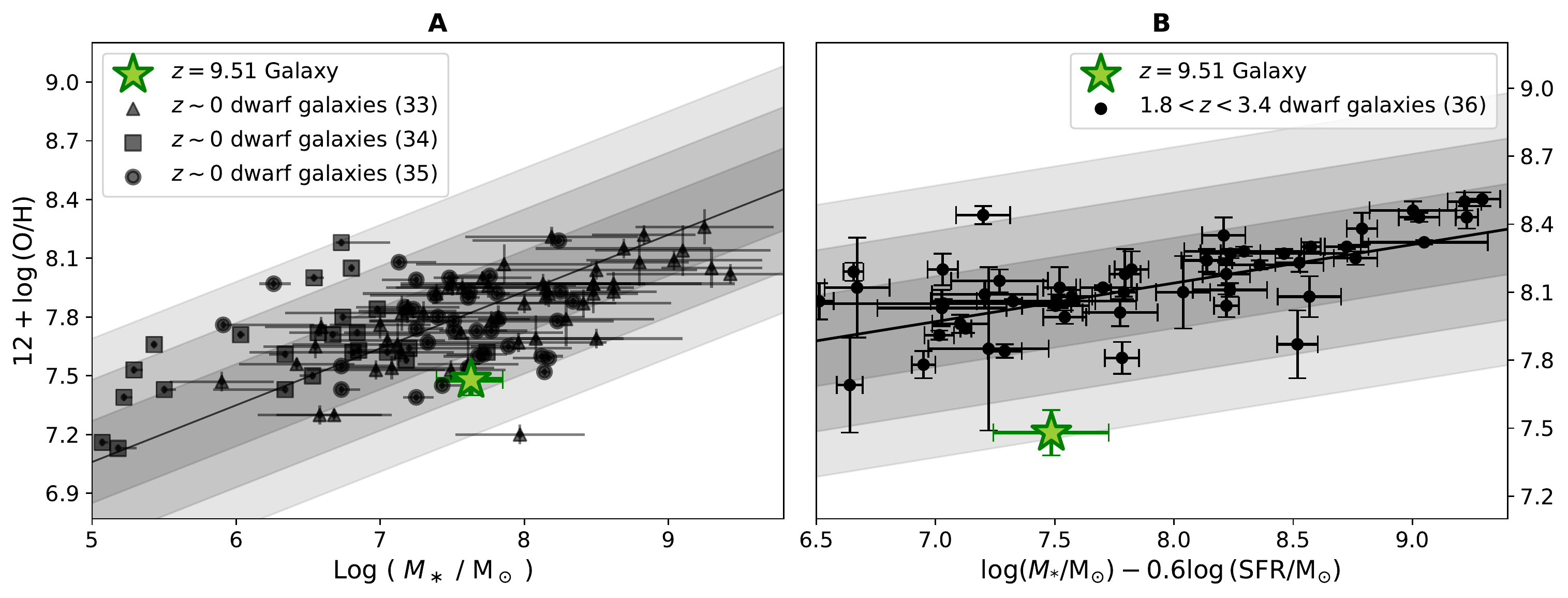}
    \caption{\textbf{Metallicity Relations. }(A) The $z=9.51$ galaxy (green star) compared to the mass-metallicity relation defined by local dwarf galaxies. Samples of local dwarfs are shown as black points \cite{Berg_2012,Hsyu_2018,Lin20XX} with error bars indicating 1$\sigma$ uncertainties. The solid line is the mass-metallicity relation fitted to the triangle data points. Gray shading indicates, from dark to light, the 1$\sigma$, 2$\sigma$, and 3$\sigma$ uncertainty ranges of this relation. (B) The more general fundamental metallicity relation (FMR) derived for dwarf galaxies at $z\sim$ 2 - 3 \cite{Li_2022}. Plotting symbols are the same. The $z=9.51$ galaxy lies 2.5$\sigma$ below this relation.}
    \label{fig:MZR}
\end{figure}

\paragraph*{Galaxy properties in context. }\label{sec:discussion}
The high magnification provided by gravitational lensing enabled us to detect this intrinsically faint galaxy ($M_{\textrm{UV},1500}=-17.4\pm0.22$ mag) with strong emission lines. Without lensing magnification, the galaxy's apparent magnitudes would be too faint to detect in the JWST images. We measured a lower mass and luminosity than other galaxies with strong emission line detections at $z > 7$, but a similar sSFR (Figure \ref{fig:highzgals}). 

Star-forming galaxies that have emission lines with very large EWs at $z\lesssim2.5$ exhibit tight correlations between the EW of the [O~\textsc{III}] 5007\ \AA \ emission line and the O$_{32}$ ratio, and between the EWs of [O~\textsc{III}] 5007 \AA \ and H$\beta$ \cite{Tang_2019}. The properties of the $z=9.51$ galaxy are consistent with both of these relations within 2$\sigma$ (Figure \ref{fig:EELG_relations}). The high $\textrm{O}_{32} = 13 \pm 4$ we measure for this object is similar to that of other galaxies with high EW emission lines at high redshifts during the epoch of reionization, and of their local counterparts \cite{Eggen2021,Flury2022}. The high O$_{32}$ might indicate a high escape fraction of hydrogen-ionizing radiation, $f_\textrm{esc}$. For example, using an empirical relation \cite{Izotov_2018}, we infer $f_{\textrm{esc}}=0.65\pm0.45$. However, there is large scatter in this relation, and other methods of inferring $f_\textrm{esc}$ do not yield such high escape fractions. For example, the UV spectral slope ($\beta = -1.98 \pm 0.11$) suggests a much smaller escape fraction, $f_{\textrm{esc}}=0.035\pm0.011$ \cite{Chisholm_2022}. Given these discrepant indicators and large uncertainties, we cannot draw any conclusions about $f_\textrm{esc}$ from this galaxy. 

The oxygen abundance is 12 + $\log$(O/H) = $7.48\pm0.08$ dex (see above), which is consistent within 2$\sigma$ with the mass-metallicity relation observed in the local Universe for similar-mass galaxies \cite{Lin20XX}. The galaxy's oxygen abundance is approximately 0.6 dex lower (2.5$\sigma$) than the more general relation between stellar mass, SFR, and metallicity (Fundamental Metallicity Relation; FMR) for dwarf galaxies at $z\sim2-3$ \cite{Li_2022} (Figure \ref{fig:MZR}). The oxygen abundances at redshift $z\gtrsim3$ are known to fall below the FMR by 0.3 - 0.6 dex \cite{Tronosco_2014}.

To determine whether the $z=9.51$ galaxy hosts an active galactic nucleus (AGN), we compare our measurements of the stellar mass and the [O~\textsc{iii}] 5007 \AA \ to H$\beta$ emission-line flux ratio ($\log$($F$([O~\textsc{iii}]) \ / $F$(H$\beta$))= 0.65 $\pm$ 0.06) to measurements from a sample of local galaxies at redshifts $0.04 < z <0.2$ \cite{Juneau_2014}. At similar stellar masses and emission line ratios to the $z=9.51$ galaxy, less than 1\% of the local galaxies were classified as AGN. If this fraction does not substantially evolve with redshift, it is unlikely that the $z=9.51$ galaxy hosts an AGN. 

\begin{figure}
    \centering
    \includegraphics[width=4.3in]{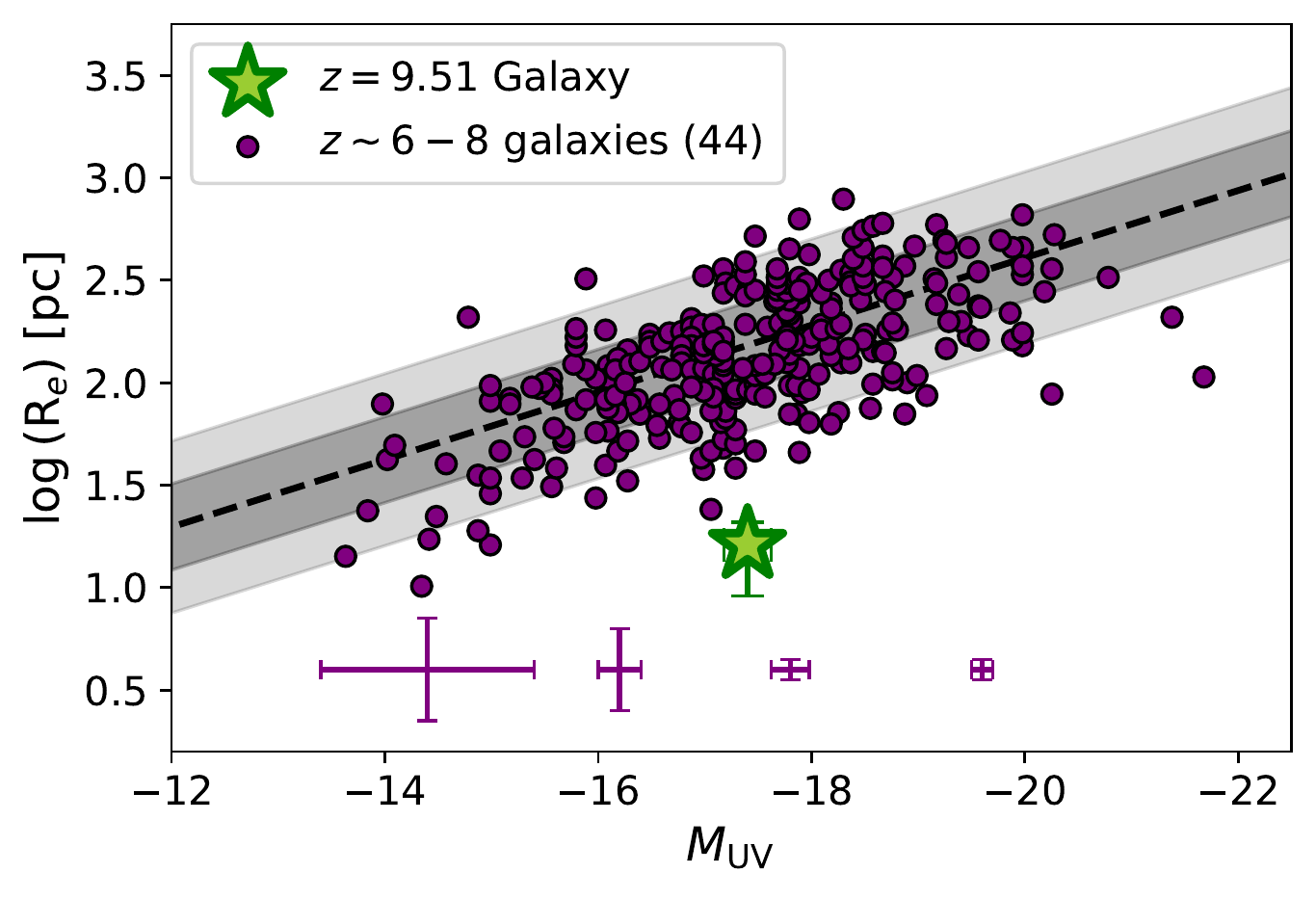}
    \caption{\textbf{Size-Luminosity Relation: } The $z=9.51$ galaxy (green star) compared to galaxies at redshifts $z\simeq 6-8$ (purple circles \cite{Bouwens_2022}). The half-light radius of the $z=9.51$ galaxy is a factor of 9.3$^{+10.5}_{-4.4}$  (3.5$\sigma$) smaller than the size-luminosity relation fitted to the purple points \cite{Bouwens_2022} (dashed line, with dark and light gray shaded regions indicating its 1$\sigma$ and 2$\sigma$ uncertainty ranges) The purple error bars indicate the typical 1$\sigma$ uncertainties for the $z\sim6-8$ galaxies.}
    \label{fig:size_comparison}
\end{figure}

We measured a half-light radius of $R_e=16.2^{+4.6}_{-7.2}$ pc for the galaxy, which is very compact compared to galaxies with similar luminosities at redshifts $z\simeq 6-8$ (Figure \ref{fig:size_comparison}). The half-light radius of the $z=9.51$ galaxy is a factor of 9.8$^{+6.5}_{-2.6}$ times smaller than the size-luminosity relation at those redshifts \cite{Bouwens_2022}, a 4$\sigma$ difference. The galaxy is also more compact than individual star-forming clumps with similar SFRs observed at redshifts $1<z<8$ \cite{Claeyssens_2023} (Figure \ref{fig:clumps}). Star-forming clumps have been shown to have a trend of increasing SFR at a fixed size with increasing redshift \cite{Livermore_2015}.

From our measurements of the SFR and half-light radius of the galaxy, we infer a very high star formation rate surface density $\Sigma_\textrm{SFR} = 1190_{-584}^{+2439}$ M$_\odot$ yr$^{-1}$ kpc$^{-2}$. $\Sigma_\textrm{SFR}$ has been observed to increase with redshift from $z \sim 0$ to $\sim 8$  \cite{Shibuya_2015} . The $\Sigma_\textrm{SFR}$ of the $z=9.51$ galaxy is a factor of 38$^{+129}_{-11}$ times higher than the galaxies in the highest redshift bin ($z\sim8$) of that sample (Figure \ref{fig:sfr_density}).

\bibliography{references}

\bibliographystyle{Science}


\noindent {\bf Acknowledgements:} 
We thank Evan Skillman, Nathan Eggen, and Alexander Criswell for very helpful comments, and Sherry Suyu for her assistance in obtaining the follow-up data.
We thank program coordinator Tricia Royle, and instrument scientists Armin Rest, Diane Karakala, and Patrick Ogle of STScI for their help carrying out the HST observations. 

{\bf Funding:} P.K. is supported by NSF grant AST-1908823 and STScI programs GO-15936, GO-16728 and GO-17253. 
M.O. acknowledges support by JSPS KAKENHI grants JP20H00181, JP20H05856, JP22H01260, and JP22K21349.
T.T. acknowledges the support of NSF grant AST-1906976.
R.J.F is supported in part by NSF grant AST--1815935, the Gordon \& Betty Moore Foundation, and by a fellowship from the David and Lucile Packard Foundation.
A.V.F. is grateful for financial assistance from the Christopher R. Redlich Fund.  G.B. is funded by the Danish National Research Foundation (DNRF) under grant \#140. 
AZ acknowledges support by Grant No. 2020750 from the United States-Israel Binational Science Foundation (BSF) and Grant No. 2109066 from the United States National Science Foundation (NSF), and by the Ministry of Science \& Technology, Israel.
J.H. and D.L. were supported by a VILLUM FONDEN Investigator grant (project number 16599).
I.P.-F. and F.P. acknowledge support from the Spanish State Research Agency (AEI) under grant number PID2019-105552RB-C43.

{\bf Author contributions:} H.W. drafted the manuscript. H.W., P.K., C.S.,  N.R., T.T., and A.V.F. revised the manuscript. W.C. reduced the spectroscopy and H.W. analysed the spectroscopy.  C.S., Y.L., N.R., T.T., W.C., D.L., A.Z, L.Y., and T.B. contributed to the interpretation. T.T., W.C., A.V.F., R.J.F., J.H., A.M.K., L.S., J.P., T.B., S.J., G.B., S.S., I.P.F., F.P., and M.N. obtained JWST imaging. G.B. measured the photometry. M.O., A.Z., and J.M.D. modeled the gravitational lensing.

{\bf Competing interests:} We declare no competing interests.

{\bf Data and materials availability:} HST imaging and JWST imaging and spectroscopy are available at \url{https://mast.stsci.edu/portal} (JWST proposal ID: 02767, HST proposal ID: 12457). Our measured photometry is provided in Table \ref{tab:phot_measurements} and line fluxes in Table \ref{tab:fluxes}. Reduced HST imaging, JWST imaging, and JWST spectroscopy is available at \url{https://zenodo.org/record/7767677} \cite{zenodo}. Subaru imaging is available at \url{http://smoka.nao.ac.jp} (Object ID: RXJ2129+0005).

\section*{Supplementary Materials}
Materials and Methods\\
Tabs. S1--S3\\
Figs. S1--S13\\
References (49--83)

\clearpage


\renewcommand{\thetable}{S\arabic{table}}
\renewcommand{\thefigure}{S\arabic{figure}}
\renewcommand{\theequation}{S\arabic{equation}}
\setcounter{figure}{0} 
\setcounter{table}{0} 
\setcounter{equation}{0}

\clearpage
  \singlespacing 
  \setcounter{page}{1}

\section*{\Large \center Supplementary Materials for}

\begin{center}
{\large \titlestring}
\end{center}

\begin{center}
Hayley Williams*, Patrick L. Kelly, Wenlei Chen, Gabriel Brammer, Adi Zitrin, Tommaso Treu, Claudia Scarlata, Anton M. Koekemoer, Masamune Oguri, Yu-Heng Lin, Jose M. Diego, Mario Nonino, Jens Hjorth, Danial Langeroodi, Tom Broadhurst, Noah Rogers, Ismael Perez-Fournon, Ryan J. Foley, Saurabh Jha, Alexei V. Filippenko, Lou Strolger, Justin Pierel, Frederick Poidevin, and Lilan Yang.
\end{center}

\begin{center}
*Corresponding author. Email: will5099@umn.edu
\end{center}

\paragraph{This PDF file includes:} 

\newenvironment{myitemize}
{ \begin{itemize}
    \setlength{\itemsep}{0pt}
    \setlength{\parskip}{0pt}
    \setlength{\parsep}{0pt}     }
{ \end{itemize}             }

\begin{myitemize}
\itemsep0em 
\setlength{\itemindent}{4.5pt}
\item[] Materials and Methods
\item[] Figs. S1-S13
\item[] Tabs. S1-S3

\end{myitemize}

\clearpage

\paragraph*{\large Materials and Methods}\label{sec:lens_modeling}

\textbf{Data Reduction: }
The JWST NIRCam observations were processed beyond the standard default calibration provided by the Space Telescope Science Institute (STScI) to produce mosaics for measuring the sources for the NIRSpec spectroscopy, as follows. After downloading the individual exposures in all the filters from the Barbara A. Mikulski Archive for Space Telescopes (MAST) portal, they were processed using version 1.8.2 of the public STScI \textsc{calwebb} software \cite{JWST_pipeline} with the photometric calibration zero points specified in the context file ``jwst\_0995.pmap". These zero points are similar to other measurements \cite{Boyer_2022}. We performed astrometric alignment between all the exposures in each filter, following previous described methods \cite{koekemoer_2011}. The absolute astrometric alignment was tied to Gaia Data Release 3 \cite{Gaia_2016,Gaia_2022} via a catalog prepared from previous SuprimeCam $z$-band imaging of the field taken for the Cluster Lensing and Supernova Survey with Hubble (CLASH) \cite{Postman_2012}. We obtained the SuprimeCam images from the Subaru Mitaka Okayama Kiso Archive. The SuprimeCam imaging was obtained on 2010 November 5 and consists of 32 exposures, each 150~s long. After applying the serial overscan correction, we performed flat-fielding and bias corrections to each of the exposures. We then performed background subtraction, following a previously published method \cite{Nonino_2009}. We used the \textsc{SWarp} software \cite{SWarp} to create the final coadded $z$-band image of the field and performed the final astrometric solution using the \textsc{SCAMP} software \cite{SCAMP}. The alignment between all the JWST exposures within each filter was achieved with a typical accuracy of $\lesssim 3$--4 milliarcseconds per source, and with $\sim 200$--500 sources per exposure this therefore achieves an overall astrometric alignment between exposures to a precision of 0.2--0.3 milliarcseconds. The final mosaics were produced at a pixel scale of 20 milliarcseconds per pixel for the F115W, F150W, and F200W images, and 40 milliarcseconds per pixel for the F277W, F356W, and F440W images.

We used the MAST portal to obtain archival HST ACS and WFC3 observations taken for CLASH. Observation dates range from 2012 April 3 to 2012 August 7 and total exposure times range from 1864~s in the F475W filter to 9530~s in the F850LP filter. The HST images were reduced using version 1.6.0 of the \textsc{grizli} software \cite{grizli}, which masks image artifacts and performs flat-fielding, background subtraction, and astrometric alignment on each exposure. The astrometric alignment is tied to Gaia Data Release 3. Reduced versions of the JWST and HST imaging are available online \cite{zenodo}. 

The NIRSpec prism data were reduced using version 1.8.5 of the public STSci \textsc{calwebb} software \cite{JWST_pipeline} and the reference files from the context file ``jwst\_1023.pmap". We downloaded the Level 1 data from the JWST archive. Detector-level corrections had already been applied to these data. We processed the Level 1 data through the \texttt{Spec2Pipeline}, which performs the background subtraction, wavelength calibration, flat-field correction, and flux calibration. The \texttt{Spec2Pipeline} also performs aperture corrections, assuming a point source target at a given position within the microshutters. Given the compact angular size of image G2 of the galaxy in the image plane ($\theta_\textrm{e, observed}=0.04\pm0.01$ arcseconds), we did not apply any additional aperture corrections. We used the {\tt Spec3Pipeline} to combine the exposures. To extract the one-dimensional (1D) spectrum, we used an optimal extraction algorithm based on a previously published method \cite{Horne_1986}. 

\begin{figure}
    \centering
    \includegraphics[width=5.0in]{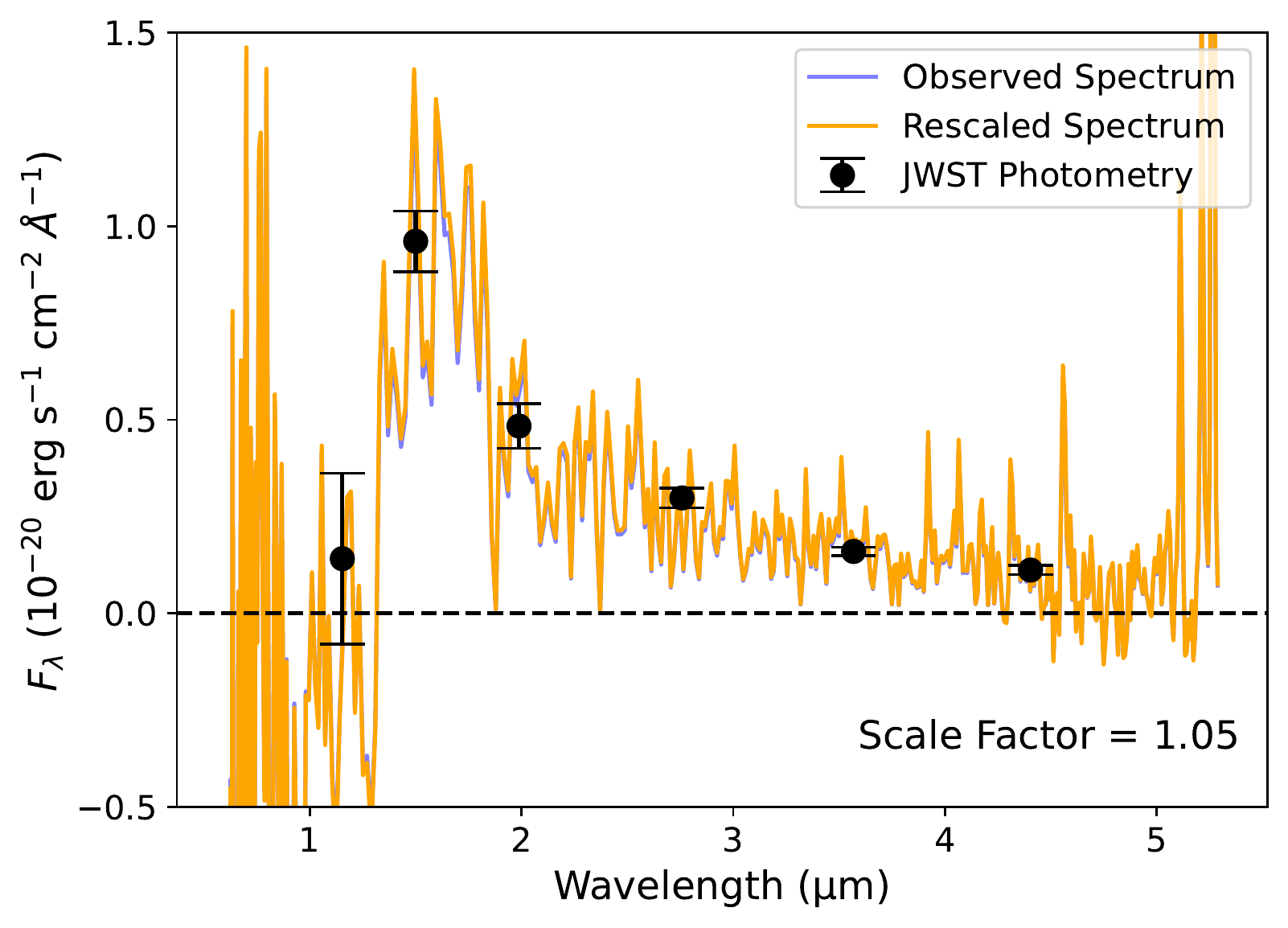}
    \caption{\textbf{Spectrum Rescaling. }Comparison between the raw and rescaled versions of the 1D extraction of the spectrum of image G2 of the $z=9.51$ galaxy. We rescale the original spectrum to match the NIRCam photometry to improve the flux calibration. The scaling factor is 1.05. Neither the spectrum nor the photometry have been corrected for magnification due to lensing.}
    \label{fig:rescale}
\end{figure}

To improve the flux calibration, we rescale our 1D spectrum to match the NIRCam wideband photometry. We compute the ratio of the mean flux density of the spectrum in 5000~\AA\ windows around the central wavelength of each NIRCam filter to the photometric flux density in that filter. We use the median raito of the six NIRCam filters as the final rescaling factor ($=1.05$). The original and rescaled version of the spectrum are shown with the NIRCam photometry in Figure \ref{fig:rescale}. The wavelength-dependent flux calibration applied in the \texttt{Spec2Pipeline} has not been verified, but the rescaled spectrum is consistent with the photometry (within $\lesssim 10\%$) for the five photometric measurements above $\sim 1.5~\upmu$m. 
Our final calibrated 1D and 2D spectra are shown in Figure \ref{fig:full_spec}.

\textbf{Imaging analysis: } To constrain the photometric redshift of the galaxy, we use the \textsc{eazy-py} software \cite{Brammer_2008}. The package compares our photometric measurements to template spectra at different redshifts and uses a $\chi^2$ likelihood function to constrain each galaxy's redshift. We obtained a photometric redshift of $z_\textrm{phot}=9.38^{+0.29}_{-0.15}$ for image G2. By fitting a linear model to the logarithmic photometric flux measurements and wavelengths in the F150W, F200W, F277W, F356W and F444W filters (spanning a rest-frame wavelength range from 950~\AA \ to 4750~\AA), we estimate a UV spectral slope ($\beta$) of -1.98 $\pm$ 0.11.

\textbf{Emission Line Flux Measurements: }
To constrain the fluxes of the emission lines, we use the \textsc{pPXF} (Penalized Pixel-Fitting) software \cite{Cappellari_2022} with the Medium-resolution Isaac Newton Telescope library of empirical spectra (MILES) \cite{Sanchez_2006, Falcon_2011}. \textsc{pPXF} uses a maximum penalized likelihood method to model the stellar continuum and fits Gaussian profiles to each of the emission lines in a given wavelength range. Since the spectral resolution of the prism data varies substantially with wavelength, we limit each run of {\tt pPXF} to $\lesssim 6000$~\AA\ windows (observer frame) around the visible emission lines. To refine our initial redshift estimate, we model the [O~\textsc{iii}] and H$\beta$ emission lines, with the line centers as a free parameter. We find $z_{\textrm{spec}}=9.51\pm0.01$. We then model each of the emission lines, with the line centers fixed to the expected wavelength of the line at redshift $z=9.51$ and line widths fixed to the spectral resolution of NIRSpec at that wavelength. Blended emission lines, such as H$\gamma$ and [O~\textsc{iii}] 4363~\AA, are modeled simultaneously and separately using two Gaussians. 

The MILES stellar library has a lower wavelength limit of 3525~\AA, so we do not use \textsc{pPXF} to model lines with rest-frame wavelengths below this limit (Ly$\alpha$ and C \textsc{iii}] 1907~\AA\ + [C \textsc{iii}] 1909~\AA). Instead, we use the nonlinear least-squares fitting routine \texttt{curve\_fit} from the \textsc{scipy} software (version 1.7.3) \cite{2020SciPy-NMeth} to fit a Gaussian profile with fixed central wavelength and width to the emission lines below 3525~\AA. Using the estimate of the continuum flux density around each emission line from \textsc{pPXF}, we also compute the rest frame EWs. 

To obtain an estimate for the upper limit of the flux of Ly$\alpha$, we inject Gaussian profiles into the spectrum at positions adjacent to the expected central wavelength of Ly$\alpha$. We repeatedly increase the flux of the injected Gaussian profile until our modeling code reports a 3$\sigma$ detection of the injected profile. We repeat this analysis at 20 positions within a 4000 \AA{} window  around the expected central wavelength of Ly$\alpha$. The average minimum flux necessary to obtain a 3$\sigma$ detection is 39 $\times 10^{-19}$ erg s$^{-1}$ cm$^{-2}$, corresponding to a rest-frame EW of approximately 31~\AA.

The observed ratio of the [O~\textsc{iii}] 4959 to [O~\textsc{iii}] 5007 emission line fluxes can be used to test the accuracy of the flux calibration in this region of the spectrum, because this ratio is set by atomic physics to be 2.98 \cite{Storey_Zeippen}. We find a flux ratio of $F$([O~\textsc{iii}]4959)/$F$([O~\textsc{iii}]5007) $=3.01\pm0.22$, consistent with this theoretical value.

Assuming Case B recombination and negligible extinction due to dust, the theoretical ratios of the fluxes of the Balmer emission lines are $F$(H$\gamma$)/$F$(H$\beta$) = 0.47 and $F$(H$\delta$)/$F$(H$\beta$) = 0.26 \cite{Osterbrock_1989}. Ratios that are smaller than these theoretical values indicate the presence of extinction due to dust, and larger ratios are unphysical. We measure larger ratios ($F$(H$\gamma$)/$F$(H$\beta$) = $0.66 \pm 0.13$ and $F$(H$\delta$)/$F$(H$\beta$) = $0.32 \pm 0.08$), but both are within 1.5$\sigma$ of the expected values for zero extinction due to dust. We proceed with our analysis assuming $A_V=0$~mag, and apply no reddening correction to the flux measurements. 

\textbf{Emission Line Analysis: } We incorporate the dust-correction uncertainty into the error analysis when calculating properties derived from intrinsic emission line fluxes and ratios.  To do so, we draw 10000 random samples from normal distributions with means set to the measured fluxes of H$\beta$ and H$\gamma$ and widths set to their 1$\sigma$ uncertainties. At each draw, we compute A$_\textrm{V}$ from the flux ratio $F$(H$\gamma$)/$F$(H$\beta$), setting A$_\textrm{V}=0$ if the flux ratio is larger than the theoretical ratio for zero extinction (see above). We find an upper 3$\sigma$ uncertainty for $A_\textrm{V}$ of 1.38~mag. 

To compute the SFR of the galaxy from the flux of the H$\beta$ emission line, we assume a previously published initial mass function (IMF) \cite{Chabrier_2003}.  The calibration we use to estimate the oxygen abundance is based on direct electron temperature ($T_{\textrm{e}}$) measurements of star forming galaxies covering O$_{32}$ values up to $\sim$ 40 \cite{Izotov2019}. 

\textbf{SED Modeling}: We use the \textsc{prospector} software \cite{Johnson_2021} to simultaneously model the NIRCam + HST photometry and the emission lines from NIRSpec. \textsc{prospector} uses the Flexible Stellar Population Synthesis package (FSPS; \cite{Conroy_2009, Conroy_2010}) to construct the stellar population models. We use \textsc{dynesty} \cite{Speagle_2020}, a dynamic nested sampling method, to explore the parameter space. Our \textsc{prospector} setup is similar to a previous study \cite{Langeroodi_2022}, the only difference is that we include the NIRSpec emission lines in the modeling. The redshift is held fixed to the spectroscopic value of $z=9.51$. We adopt a non-parametric model for the SFH, consisting of five independent temporal bins: the most recent bin spans 0-10 Myr in lookback time, and the remaining four bins are evenly spaced in $\log$(lookback time) up to $z=35$, which is the earliest onset of star formation allowed in the \textsc{FSPS} code. The scaling factor for the optical depth of the intergalactic medium (IGM) $f_{\textrm{IGM}}$ is a free parameter in the model, for which we use a redshift-dependent IGM attenuation model \cite{Madau_1995}.

We assume an IMF \cite{Chabrier_2003} and a dust attenuation curve \cite{Kriek_2013}. The dust attenuation is modeled with three free parameters: the optical depth at 5500$\textrm{\AA}$ ($\tau_\textrm{V}$), the dust index $\Gamma_{\textrm{dust}}$ which controls the strength of the UV bump in the attenuation curve, and the dust ratio $r_{\textrm{dust}}$ which determines the ratio of the optical depth affecting young stars compared to the optical depth affecting all stars. Additional free parameters in our model include the stellar metallicity $\log$($Z_*$/$Z_\odot$) where $Z_\odot$ is the solar metallicity, the nebular metallicity $\log$($Z_{\textrm{neb}}$/$Z_\odot$), and the nebular ionization parameter $\log$($U_{\textrm{neb}}$). The model fits for the total formed stellar mass ($\log$($M_{\textrm{*,formed}}$/$M_\odot$)), which is the integral of the SFH and does not account for stellar mass loss during evolution due to supernovae and winds. The total formed stellar mass can be converted to the current stellar mass ($\log$($M_{\textrm{*}}$/$M_\odot$)) at each step in the parameter space. The priors and allowed ranges for each of the free parameters in our model fitting are shown in Table \ref{tab:prospector}.

The emission lines from our NIRSpec data are included in the \textsc{prospector} modeling. The line widths are fixed to the instrumental resolution of the NIRSpec prism at their observed wavelengths. We only include emission lines above 3.7 $\mu$m, a limit set by the availability of sufficiently high resolution libraries of stellar spectra used by {\tt prospector} and {\tt FSPS} (MILES library \cite{Sanchez_2006} which spans the rest-frame 3525 to 7400 \AA). This wavelength range includes all of the emission lines we measured with S/N $>3$ (Table \ref{tab:fluxes}). To account for potential uncertainties in spectroscopic calibration, following \cite{Johnson_2021}, at each likelihood call we multiply the model spectrum by a maximum-likelihood 12th order polynomial that describes the ratio between the model and observed spectrum.

\begin{figure*}
    \centering
    \includegraphics[width=\textwidth]{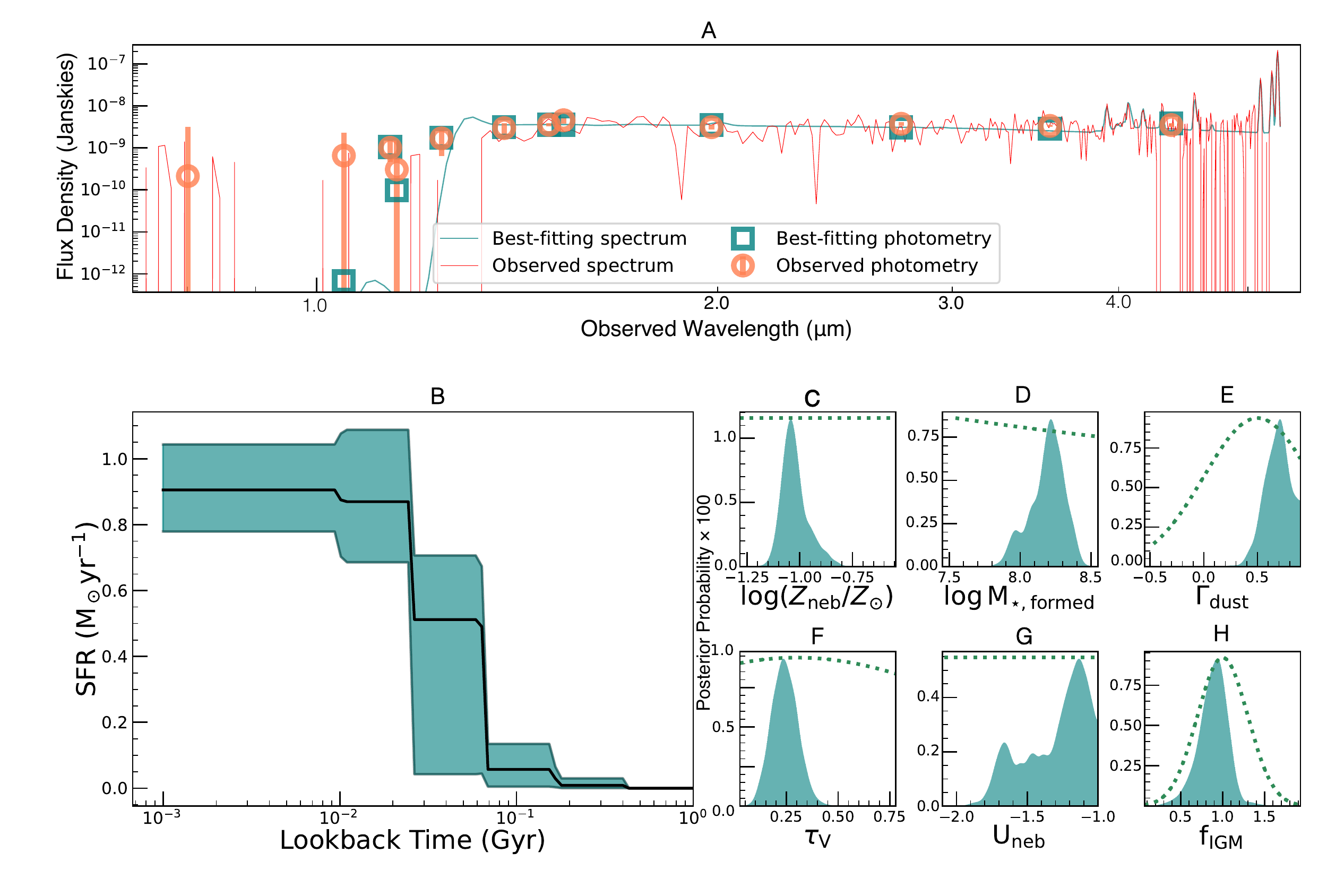}
    \caption{\textbf{Best-Fitting SED model: } (A) Observed NIRCam and \textit{HST} photometry of image G2 of the $z=9.51$ galaxy (orange circles), with the best-fitting model from \textsc{prospector} (teal squares). The portion of the NIRSpec spectrum used in the SED fitting ($>3.7 \mu$m) is shown in red, and the best-fitting spectrum from \textsc{prospector} is shown as the teal line. The observed and best-fitting spectra and photometry have been corrected for magnification due to lensing. (B) The best-fitting SFH from \textsc{prospector}. The black line shows the best-fitting SFR as a function of lookback time and the teal shading shows the 1$\sigma$ uncertainty ranges in each temporal bin. (C-H) The posterior probability distribution functions (PDFs, shown as solid teal regions) for six free parameters. The dotted teal lines above the PDFs show the priors used in the fitting. (C) PDF of the nebular metallicity. (D) PDF of the total formed stellar mass. (D) PDF of the dust index. (E) PDF of the optical depth at 5500~\AA. (F) PDF of the ionization parameter. (G) PDF of the scaling factor for the optical depth of the IGM.}
    \label{fig:photometry_11027}
\end{figure*}

Figure \ref{fig:photometry_11027} compares the best-fitting spectrum to the observed SED of image G2, along with the posterior probability distributions for a selection of free parameters. To test for systematic bias, we run \textsc{prospector} using the same parameters but with different combinations of the observed photometry and spectroscopy (full spectroscopy + photometry, emission lines + photometry, photometry only) and with differing lower wavelength limits for the included spectroscopy. The results are consistent for the majority of the parameters, but we add additional systematic uncertainties (in quadrature) of 0.2 dex in the stellar mass, 0.15 magnitudes in the extinction due to dust, and 0.3 M$_\odot$ yr$^{-1}$ in the SFR to reflect the range of best-fitting values. 


\textbf{Lens Modeling: } We construct a mass model of the cluster RX\,J2129 using the \textsc{glafic} software \cite{Oguri_2010,Oguri_2021}. A previous \textsc{glafic} mass model for this cluster has been derived \cite{Okabe_2020}, in which 22 multiple images of 7 background galaxies \cite{Zitrin_2015,Richard2010MNRAS.404..325R,Caminha_2019} were used as constraints. Spectroscopic redshifts of these 7 galaxies have also been measured  \cite{Caminha_2019}. Our mass model consist of a single elliptical dark matter \cite{Navarro_1997} halo component representing the cluster, with member galaxies modeled by an ellipsoid profile \cite{Keeton_2001} with their velocity dispersions and truncation radii scaled with their luminosities \cite{Kawamata_2016}. This mass model reproduces the counterimages of G2, given its location and spectroscopic redshift of $z=9.51$. By searching for objects that are located near the predicted counterimage positions and have colors similar to those of G2, we identify the counterimage candidates G1 and G3. We then refine our mass model by adding the positions of G1, G2, and G3 as constraints. The resulting best-fitting mass model reproduces all the positions of multiple images including G1--G3, with the root-mean square of the positional differences between observed and predicted image positions of $0.52''$. To estimate the uncertainties, we run a Markov chain Monte Carlo (MCMC) process for the mass model, assuming the positional error of $0.4''$. We use this refined mass model to calculate the magnification of G1 to $10.5 \pm 1.7$, that of G2 is $20.2 \pm 3.8$, and that of G3 is $11.3 \pm 1.7$.

We model the photometry of the identified counterimages using \textsc{prospector}. The fit parameters and their respective priors are identical to those used for image G2. There is no available spectroscopy for the counterimages, so we use only the observed HST and NIRCam photometry (corrected for magnification).  Because the magnification of the counterimages is a factor of $\sim2$ smaller than that of image G2, the photometry is less constraining and the best-fitting parameters from the modeling have higher uncertainties. For image G1, we infer $\log$(M$_*$/M$_\odot$) = $8.03^{+0.36}_{-0.43}$, and for image G3, we infer $\log$(M$_*$/M$_\odot$) = $8.34^{+0.43}_{-0.45}$. These values are larger than what we inferred for image G2 ($\log$(M$_*$/M$_\odot$) = $7.63^{+0.22}_{-0.24}$), differing by $\sim$1.5$\sigma$. 

To verify these results, we also use a second mass model constructed using an analytic version of the Zitrin-parametric code \cite{Zitrin_2015}. This version is not coupled to a fixed grid resolution and thus capable of high resolution results (similar or better than the spatial resolution of the JWST images). We use the same set of constraints as above, including the $z=9.51$ counterimages identified by the \textsc{glafic} model. Galaxies are modeled as double pseudo isothermal elliptical density profiles, based on common scaling relations, and the dark matter halo as a pseudo isothermal elliptical mass distribution \cite{Furtak_2022}. The minimization is performed in the source plane using a several thousand step MCMC process. For the minimization and uncertainty estimation we use a positional uncertainty of $0.5''$ for most multiple images, but adopt $0.2''$ for the three images of the $z=9.51$ galaxy and the system in which a multiply-imaged supernova was seen. We infer magnifications and 95\% confidence intervals of $8.7^{+1.6}_{-4.3}$, $18.6^{+5.8}_{-8.4}$, and $9.9^{+2.9}_{-3.4}$ for G1, G2, and G3, respectively, which are consistent with the \textsc{glafic} model within the 1$\sigma$ uncertainties.

\textbf{Cosmology: }Throughout this paper, we assume a flat lambda-cold dark matter cosmology with matter density parameter $\Omega_{m}=0.287$ and Hubble parameter H$_0=69.3$~km~s$^{-1}$~Megaparsec$^{-1}$. All dates are reported in universal time.



\begin{table}[ht]
\centering
\caption{\textbf{JWST and HST Photometry: }Photometry measurements of image G2 of the $z=9.51$ galaxy from the JWST NIRCam, HST WFC3, and HST ACS imaging. The naming convention for the HST filters is the same as that of the JWST filters. These measurements have not been corrected for magnification due to lensing. Uncertainties are 1$\sigma$. Upper limits are 3$\sigma$.}
\setlength{\tabcolsep}{15pt}
\begin{tabular}{@{}ccc@{}}
\\
\multicolumn{3}{ c }{\textbf{JWST NIRCam}} \\
\midrule
\textbf{Filter Name} & \textbf{Wavelength (\AA)} & \textbf{Flux Density (10$^{-20}$ erg s$^{-1}$ cm$^{-2}$ \AA$^{-1}$)}\\
\midrule
F115W & 11543 & 0.14 ± 0.22\\
F150W & 15007 & 0.96 ± 0.08\\
F200W & 19886 & 0.48 ± 0.06\\
F277W & 27578 & 0.30 ± 0.03\\
F356W & 35682 & 0.16 ± 0.01\\
F440W & 44037 & 0.11 ± 0.01\\
\midrule
\\
\multicolumn{3}{ c }{\textbf{HST WFC3}} \\
\midrule
\textbf{Filter Name} & \textbf{Wavelength (\AA)} & \textbf{Flux Density (10$^{-20}$ erg s$^{-1}$ cm$^{-2}$ \AA$^{-1}$)}\\
\midrule
F225W & 2370 & 31.81 ± 27.64\\ 
F336W & 3354 & $<25.02$\\ 
F606W & 5892 & $<3.00$ \\ 
F105W & 10544 & 0.36 ± 0.45\\ 
F110W & 11534 & 0.46 ± 0.20\\ 
F125W & 12471 & 0.66 ± 0.32\\ 
F140W & 13924 & 0.93 ± 0.25\\ 
F160W & 15397 & 1.17 ± 0.19\\ 
\midrule
\\
\multicolumn{3}{ c }{\textbf{HST ACS}} \\
\midrule
\textbf{Filter Name} & \textbf{Wavelength (\AA)} & \textbf{Flux Density (10$^{-20}$ erg s$^{-1}$ cm$^{-2}$ \AA$^{-1}$)}\\
\midrule
F435W & 4319 & 3.54 ± 21.48\\ 
F475W & 4747 & 1.18 ± 6.15\\ 
F606W & 5921 & $<3.24$\\ 
F625W & 6311 & $<7.11$ \\ 
F775W & 7692 & $<22.35$\\ 
F814W & 8057 & 0.20 ± 0.54\\ 
F850LP & 9033 &$<2.40$\\
\midrule
\end{tabular}

\label{tab:phot_measurements}
\end{table}

\begin{table}[ht]
\centering
\caption{\textbf{Coordinates of the images of RX\,J2129-z95. }Right ascension and declination for the three images (G1, G2, and G3) of the $z=9.51$ galaxy.}
\setlength{\tabcolsep}{15pt}
\begin{tabular}{@{}ccc@{}}
\\
\textbf{Image} & \textbf{Right Ascension (hh:mm:ss)} & \textbf{Declination (dd:mm:ss)}\\
\midrule
RX\,J2129-z95:G1 & 21:29:40.858 & +00:05:37.096\\
RX\,J2129-z95:G2 & 21:29:41.175 & +00:05:30.075 \\
RX\,J2129-z95:G3 & 21:29:41.381 & +00:05:20.166 \\
\end{tabular}
\label{tab:coordinates}
\end{table}

\begin{table}[ht]
\centering
\caption{\textbf{SED modeling parameters: }The free parameters used in \textsc{prospector} for fitting the model SED, with their input priors, allowed ranges, and the output best-fitting values. $\log$(M$_{*}$/M$_\odot$) is the stellar mass (the model fits for the total mass formed but we convert here to account for stellar mass loss during evolution). $\log(Z_*/Z_{\odot})$ and $\log(Z_\textrm{neb}/Z_{\odot})$ are the stellar and nebular metallicities, respectively. $\log(\textrm{U}_{\textrm{neb}})$ is the logarithm of the nebular ionization parameter. $f_\textrm{IGM}$ is the scaling factor for the optical depth of the IGM absorption. $\tau_\textrm{V}$, $\Gamma_\textrm{dust}$, and $r_\textrm{dust}$ are the dust attenuation parameters: they control the optical depth at 5500~\AA, the strength of the UV bump in the dust attenuation curve, and the ratio of the optical depth affecting young stars compared to the optical depth affecting all stars, respectively.}
\setlength{\tabcolsep}{14pt}
\begin{tabular}{@{}lllc@{}}
\\
\textbf{Parameter}& \textbf{Prior} & \textbf{Allowed Range} & \textbf{Best-Fitting Value} \\
\midrule
$\log$(M$_{*}$/M$_\odot$)& Log Uniform &[6, 10] & 7.63$^{+0.10}_{-0.14}$\\

$\log(Z_*/Z_{\odot})$ & Top Hat & [-2, 0.19] & -1.70$^{+0.30}_{-0.18}$ \\
$\log(\textrm{Z}_{\textrm{neb}}/\textrm{Z}_{*})$ & Top Hat & [-2, 0.5] & -1.04$^{+0.08}_{-0.05}$ \\
$\log(\textrm{U}_{\textrm{neb}})$& Top Hat & [-4, -1] & -1.24$^{+0.15}_{-0.37}$\\
$f_{\textrm{IGM}}$ & Normal &[0, 2]& 0.90$^{+0.15}_{-0.18}$ \\
 & (mean = 1.0, width = 0.3) & & \\

$\tau_{\textrm{V}}$ & Normal &[0, 4] & 0.24$^{+0.06}_{-0.06}$\\
 & (mean = 0.3, width = 1.0) & & \\
$\Gamma_{\textrm{dust}}$&  Normal &[-1, 0.4] & 0.19$^{+0.11}_{-0.13}$\\
 & (mean = 0.0, width = 0.5) & & \\
r$_{\textrm{dust}}$ & Normal &[0, 1.5] & 1.03$^{+0.15}_{-0.15}$\\
 & (mean = 1.0, width = 0.3) & & \\
\end{tabular}

\label{tab:prospector}
\end{table}


\begin{figure}[ht]
     \centering
     \includegraphics[width=\textwidth]{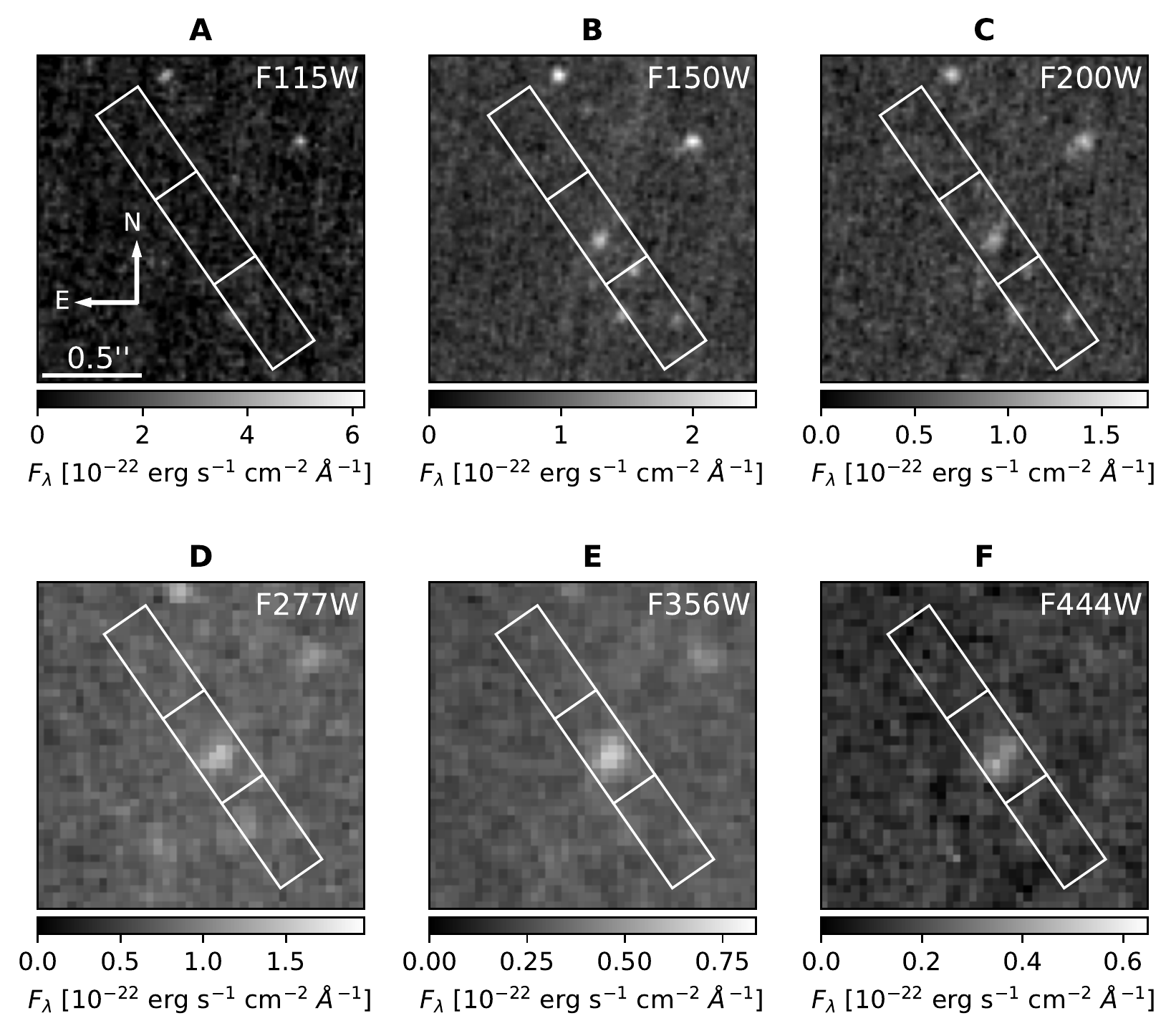}
     \label{fig:11027_img}
    \caption{\textbf{NIRCam Infrared Images of the $z=9.51$ galaxy. } Image G2 of the $z=9.51$ galaxy in six NIRCam filters. Each panel (A-F) is labeled with the name of the NIRCam filter. The J2000 position of image G2 is ($\alpha,~\delta$) = (21$^{\rm h}$29$^{\rm m}$41.173$^{\rm s}$, $+00^\circ05'30.073''$). The white boxes show the positions of the open NIRSpec microshutters used for the observation. The scale and orientation shown in panel A are the same for all panels. (A) F115W NIRCam image. (B) F150W NIRCam image. (C) F200W NIRCam image. (D) F277W NIRCam image. (E) F356W NIRCam image. (F) F444W NIRCam image.}
    \label{fig:img}
\end{figure}

\begin{figure}
    \centering
    \includegraphics[width=\textwidth]{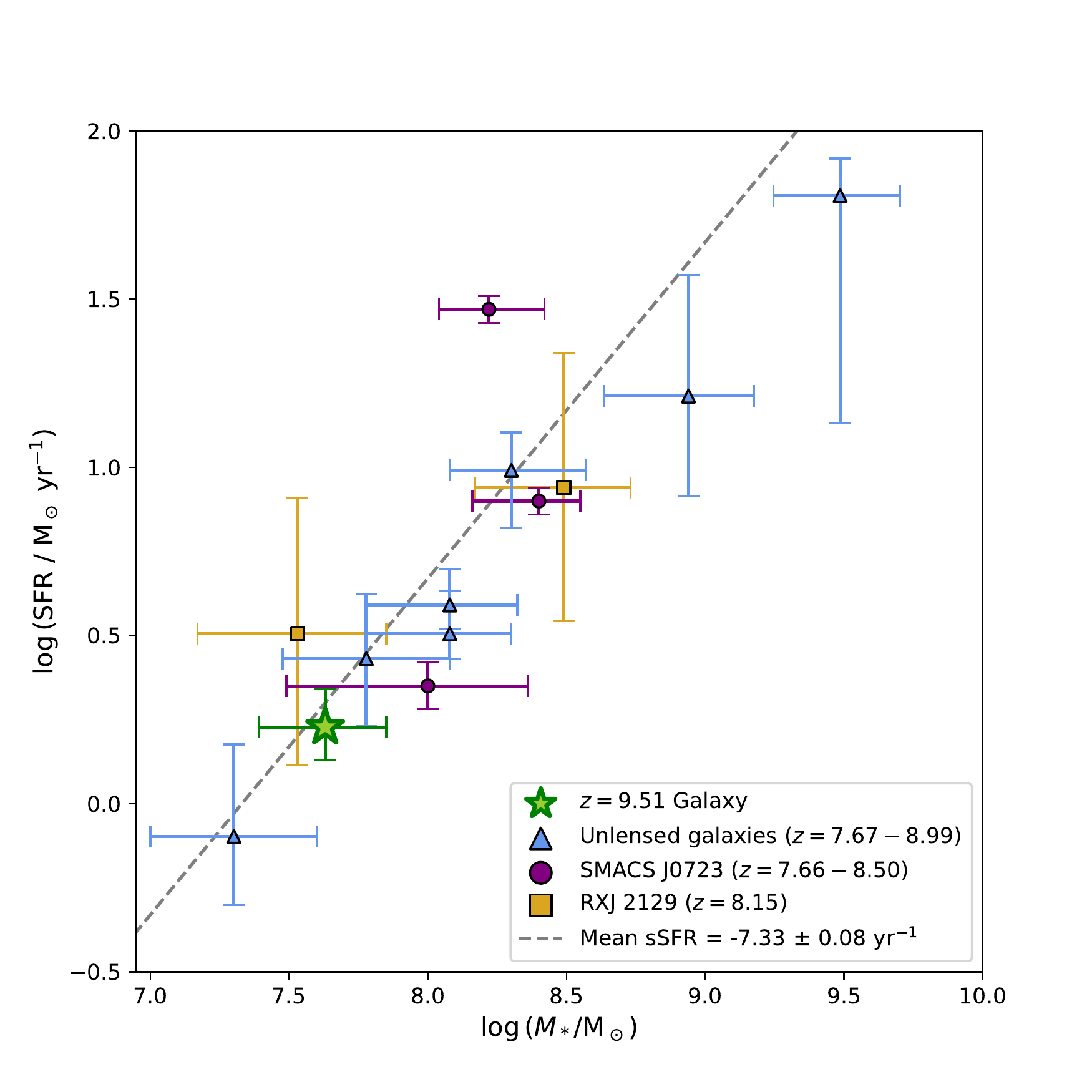}
    \caption{\textbf{Spectroscopically Confirmed $z>7$ Galaxies: }The stellar mass and SFR of our $z=9.51$ galaxy (green star) compared to other spectroscopically confirmed galaxies at $7<z<10$. The mean sSFR of the sample is -7.33 $\pm$ 0.08 yr$^{-1}$ (dashed gray line), which is very similar to the value measured for the $z=9.51$ galaxy (sSFR = -7.38 $\pm$ 0.26). The unlensed galaxies \cite{Fujimoto_2023} are shown as blue triangles. The SMACS\,J0723 galaxies (masses from \cite{Langeroodi_2022} and SFRs from \cite{Heintz_2022}) are shown as purple circles. The RX\,J2129 $z\sim8$ galaxies \cite{Langeroodi_2022} are shown as yellow squares. Error bars show 1$\sigma$ uncertainties. The stellar masses and SFRs have been corrected for magnification due to lensing.}
    \label{fig:highzgals}
\end{figure}

\begin{figure}
    \centering
    \includegraphics[width=\linewidth]{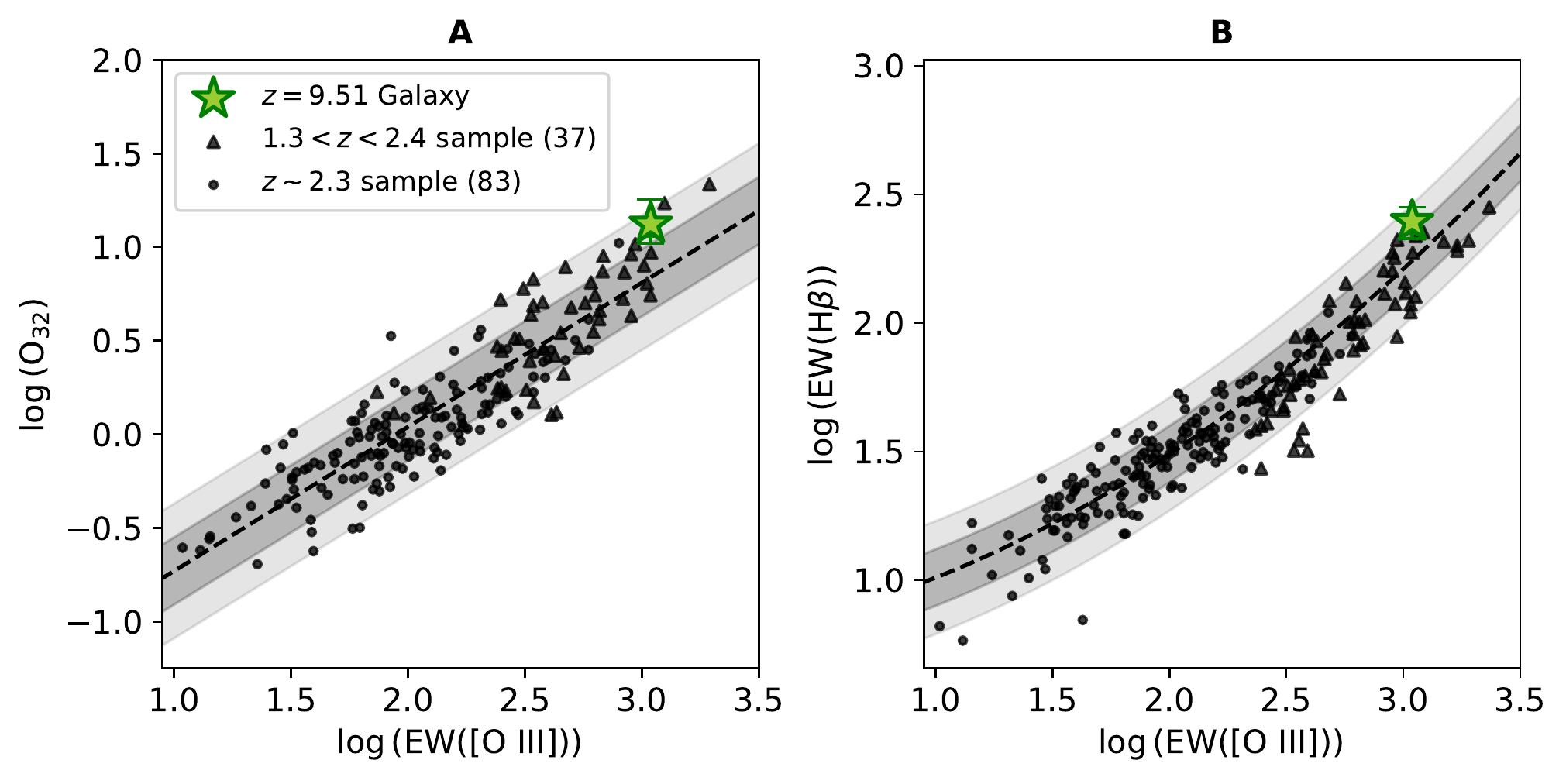}
    \caption{\textbf{Extreme Emission Line Galaxy Relations: }The $z=9.51$ galaxy (green star) compared with the  emission line relations for extreme emission line galaxies (EELGs)at $z \lesssim 2.5$ \cite{Sanders_2020}. Panel A shows the relation between the EW of [O~\textsc{iii}]$\lambda$5007 and $\textrm{O}_{32}$. Panel B shows the relation between the EW of [O~\textsc{iii}]$\lambda$5007 and the EW of H$\beta$ . A sample of EELGs at $1.3 < z <2.4$ \cite{Tang_2019} are shown as black triangles and a sample of star-forming galaxies at $z\sim2.3$ \cite{Sanders_2018} are shown as black circles. The 1$\sigma$ and 2$\sigma$ ranges for the EELG relations are shown in dark and light gray, respectively.}
    \label{fig:EELG_relations}
\end{figure}

\begin{figure}
    \centering
    \includegraphics[width=0.6\linewidth]{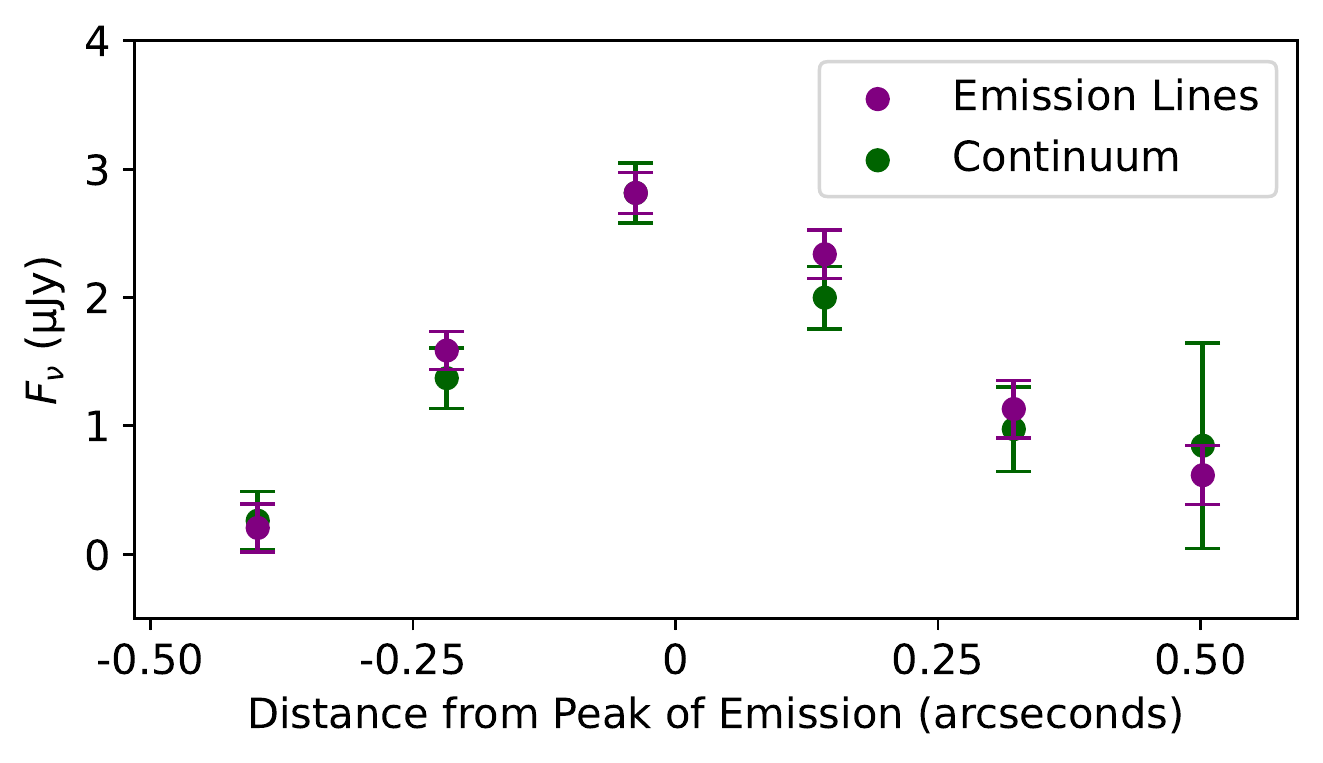}
    \caption{\textbf{Spatial Emission Profiles: } We test for an offset between the nebular emission and the stellar continuum of the $z=9.51$ galaxy by extracting profiles along the spatial axis of the NIRSpec MSA slit. The emission line spatial profiles are shown in purple, and the stellar continuum spatial profile is shown in green. There is no evidence for an offset between the nebular emission and the stellar continuum.}
    \label{fig:profiles}
\end{figure}

\begin{figure}
    \centering
    \includegraphics[width=\linewidth]{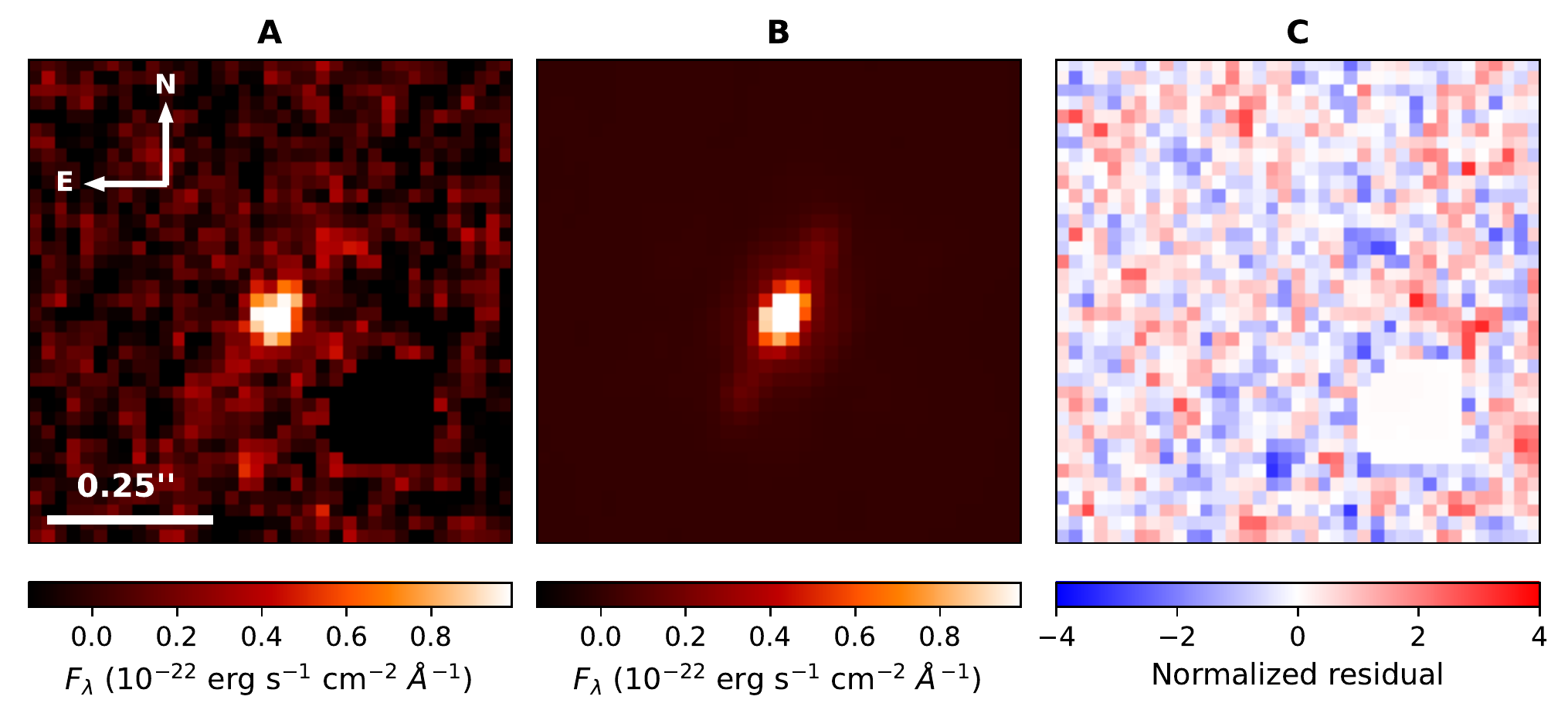}

    \caption{\textbf{Galaxy Size Modeling. }Modeling of the size of image G2 of the $z=9.51$ galaxy. Panel A shows the observed NIRCam F150W image G2. Panel B shows the modeled S\`{e}rsic profile of image G2. Panel C shows the normalized residuals between the model profile and the observed data (observed flux - model flux / standard deviation of observed flux). The scale and orientation shown in panel A are the same for all panels. We measure an intrinsic half-light radius in the source plane of $16.2^{+4.6}_{-7.2}$~pc.} 
    \label{fig:imfit}
\end{figure}

\begin{figure}
    \centering
    \includegraphics{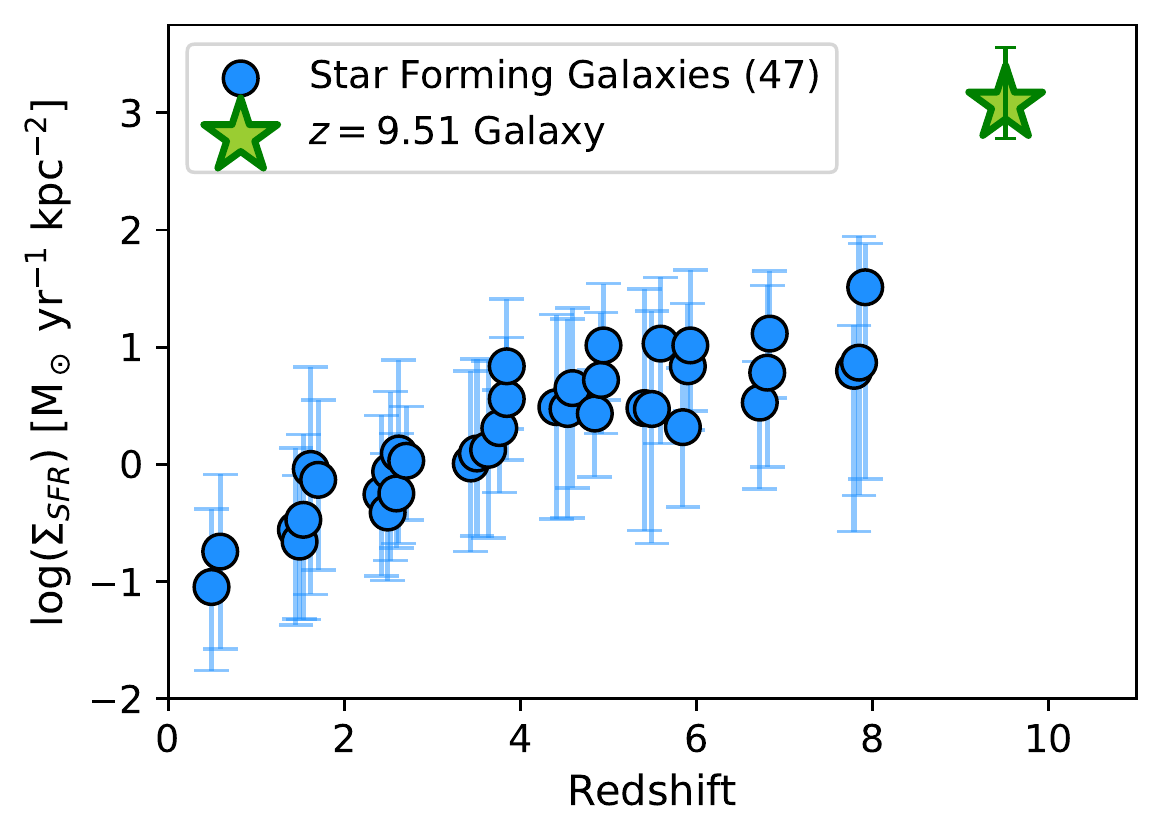}
    \caption{\textbf{Star Formation Rate Surface Density: } The star formation rate surface density $\Sigma_\textrm{SFR}$ of the $z=9.51$ galaxy compared to star forming galaxies (SFGs) spanning a redshift range $0<z<10$. The blue points show redshift-binned stacks of $\Sigma_\textrm{SFR}$ measurements of a total of $\sim$ 190,000 galaxies \cite{Shibuya_2015}. The $\Sigma_\textrm{SFR}$ of the $z=9.51$ galaxy is a factor of 38$^{+129}_{-11}$ larger the SFGs in the highest redshift bin ($z\sim8$). Error bars show 1$\sigma$ uncertainties.}
    \label{fig:sfr_density}
\end{figure}

\begin{figure}
    \centering
    \includegraphics[width=\linewidth]{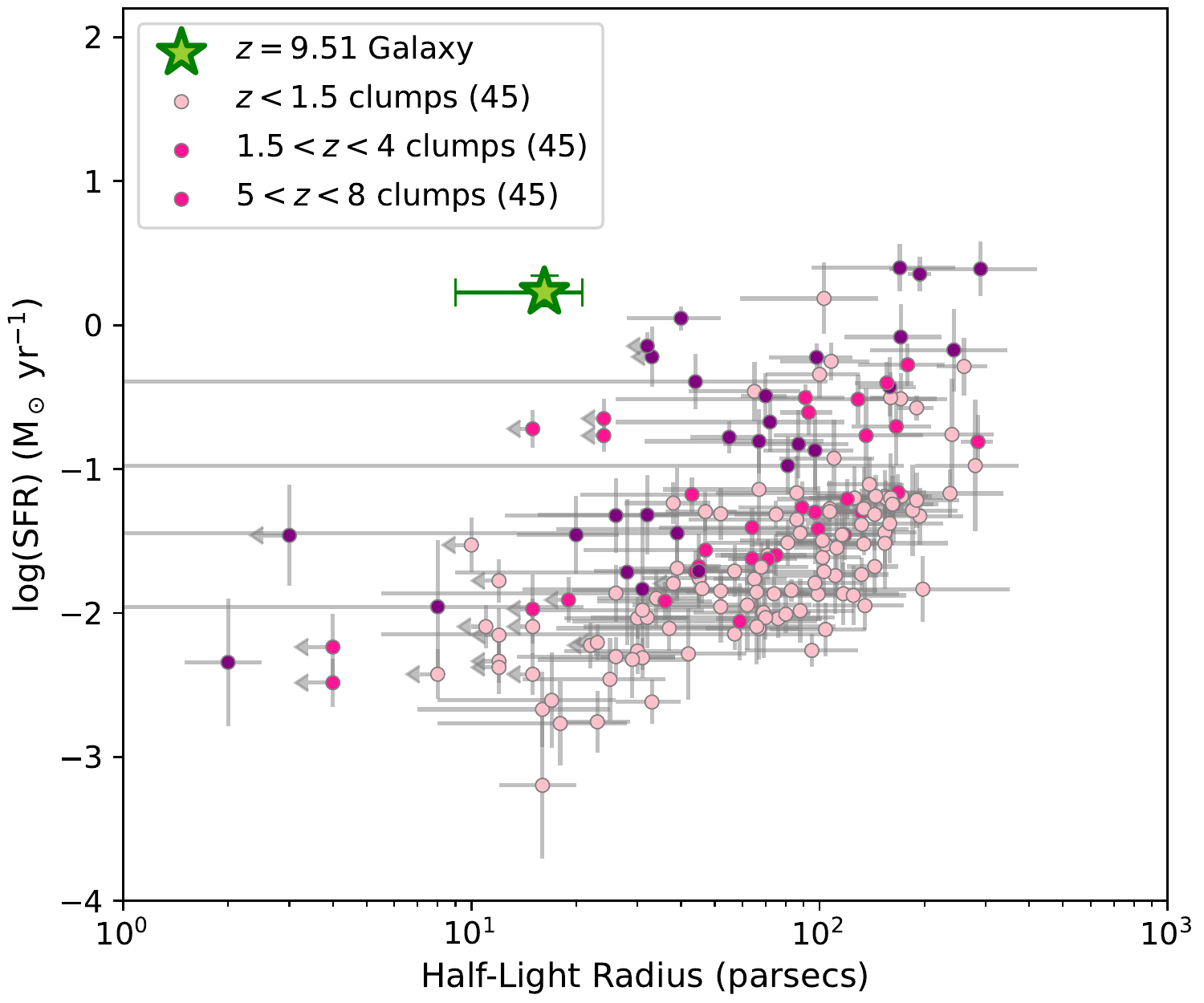}
    \caption{\textbf{Comparison With Star-Forming Clumps. }The $z=9.5$ galaxy (green star) compared to a sample of gravitationally lensed star-forming clumps in the the SMACS 0723 cluster field at $1<z<8$ (pink and purple points) \cite{Claeyssens_2023}. There is a trend of increasing SFR at a fixed radius with increasing redshift \cite{Livermore_2015}.  The SFR of the $z=9.51$ galaxy has been corrected for magnification due to lensing. Error bars show 1$\sigma$ uncertainties.}
    \label{fig:clumps}
\end{figure}

\begin{figure*}
    \centering
    \includegraphics[width=\linewidth]{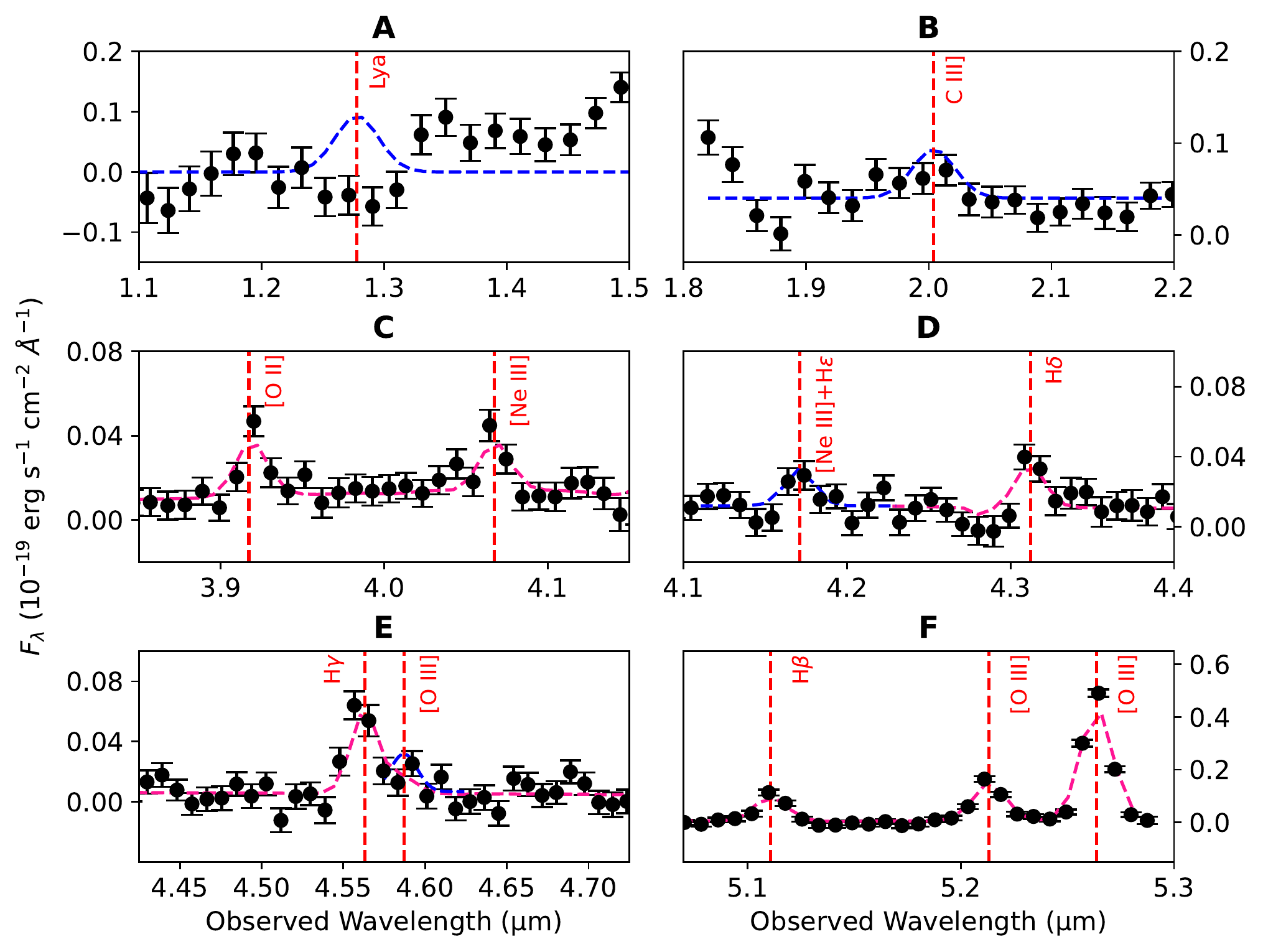}
    \caption{\textbf{Emission Line Profiles. } Detected emission lines from image G2 of the $z=9.51$ galaxy, with Gaussian models fitted to the emission-line profiles using \textsc{pPXF} shown in pink. 3$\sigma$ upper limits of the undetected lines are shown in blue. The red dashed lines show the expected observed wavelength of each emission line at $z=9.51$. Error bars show 1$\sigma$ uncertainties. Emission-line flux measurements are given in Table \ref{tab:fluxes}. These spectra have not been corrected for magnification due to lensing. }
    \label{fig:11027_ppxf}
\end{figure*}

\begin{figure*}
\centering
\includegraphics[width=\linewidth]{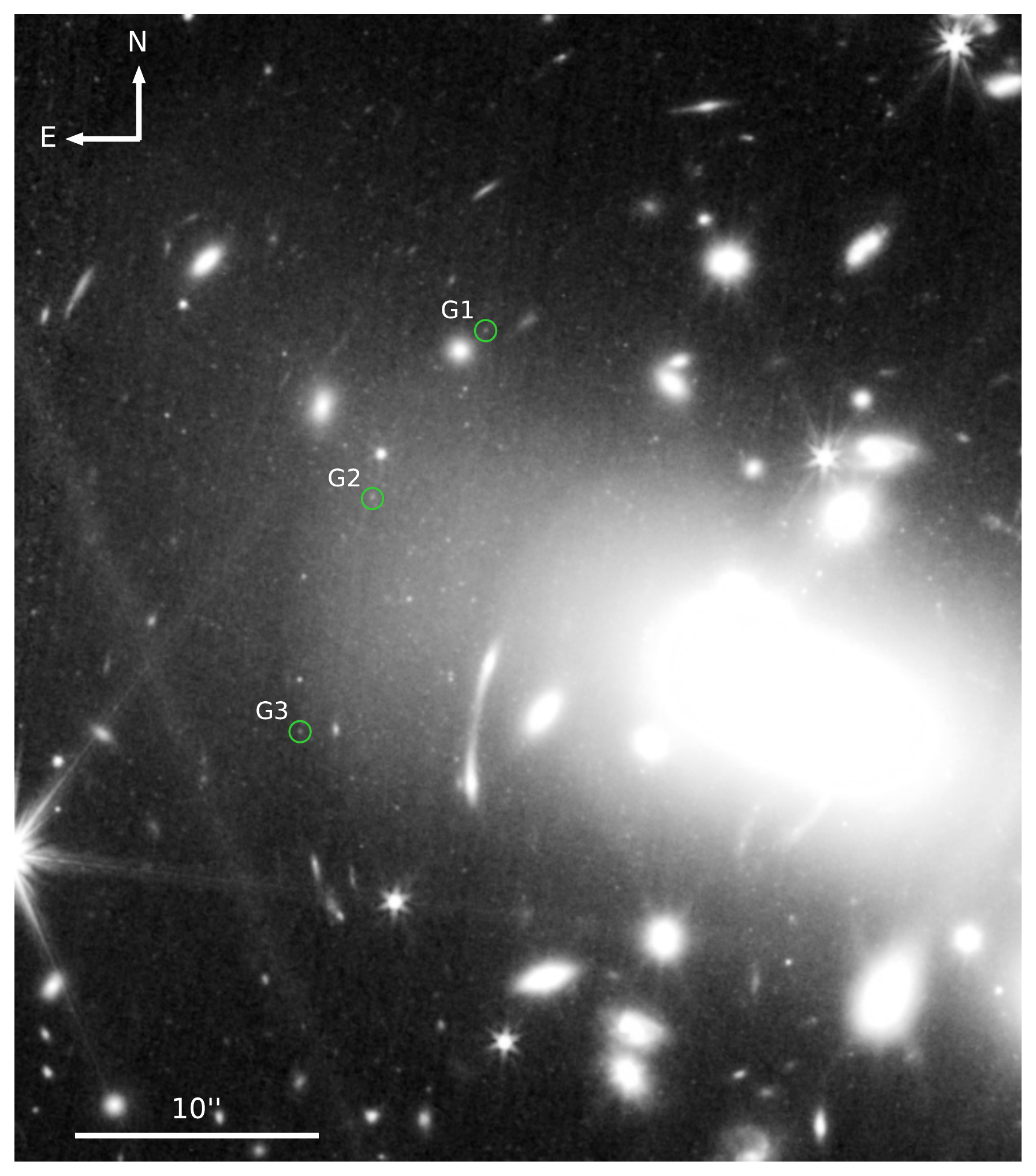}
\caption{\textbf{Red frame of RX\,J2129 color-composite image. } The red frame from the color-composite image of part of RX\,J2129 (Figure \ref{fig:imglarge}), with three images of the $z=9.51$ galaxy circled in green. This image is a composite of the JWST NIRCam F277W + F356W + F444W images. The broad bands are diffraction spikes from foreground stars.}
\label{fig:rxj2129_red}
\end{figure*}

\begin{figure*}
\centering
\includegraphics[width=\linewidth]{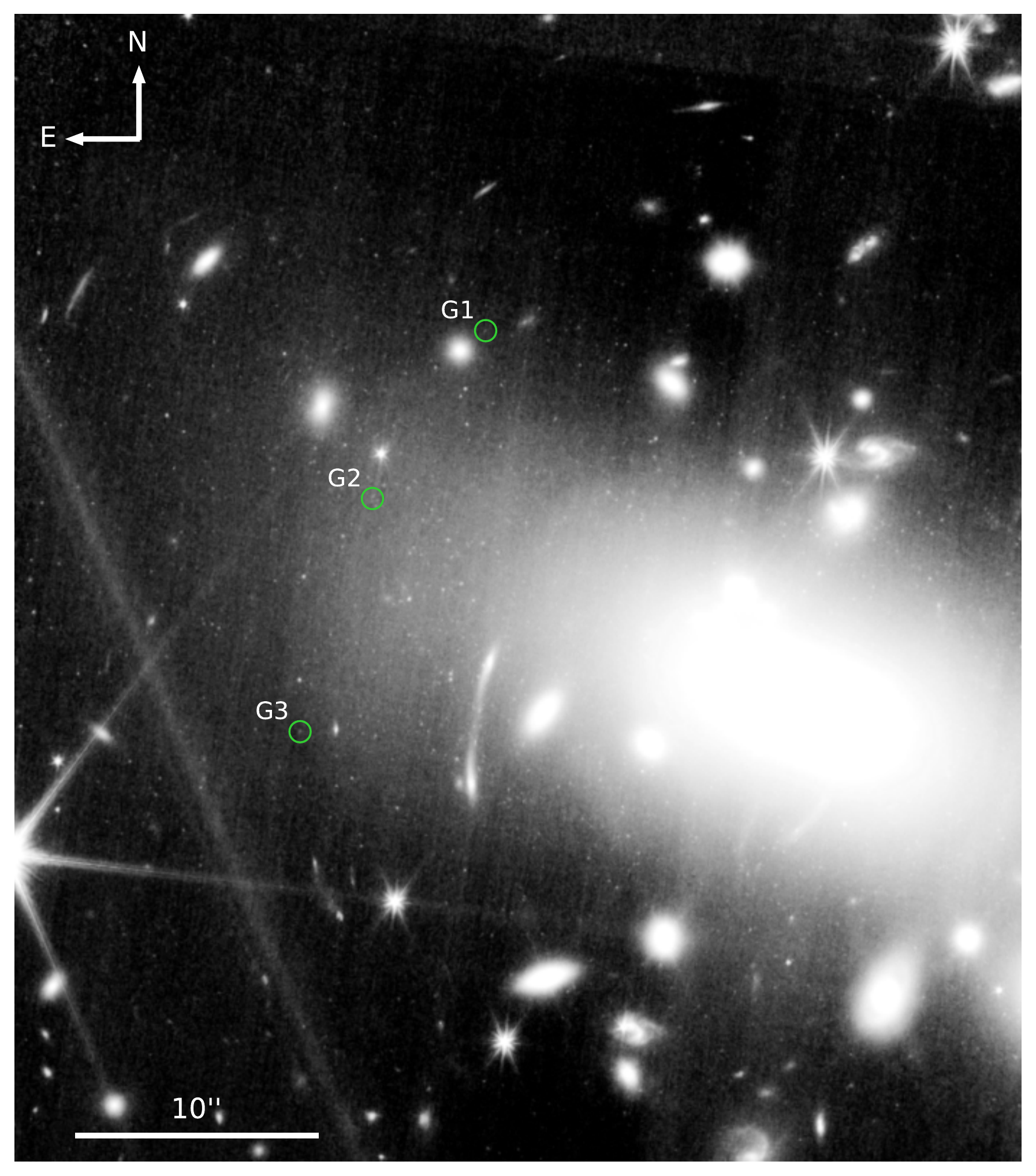}
\caption{\textbf{Green image of part of RX\,J2129. } The green frame from the color-composite image of part of RX\,J2129 (Figure \ref{fig:imglarge}), with three images of the $z=9.51$ galaxy circled in green. This image is a composite of the JWST NIRCam F115W + F150W + F200W images. The broad bands are diffraction spikes from foreground stars.}
\label{fig:rxj2129_green}
\end{figure*}

\begin{figure*}
\centering
\includegraphics[width=\linewidth]{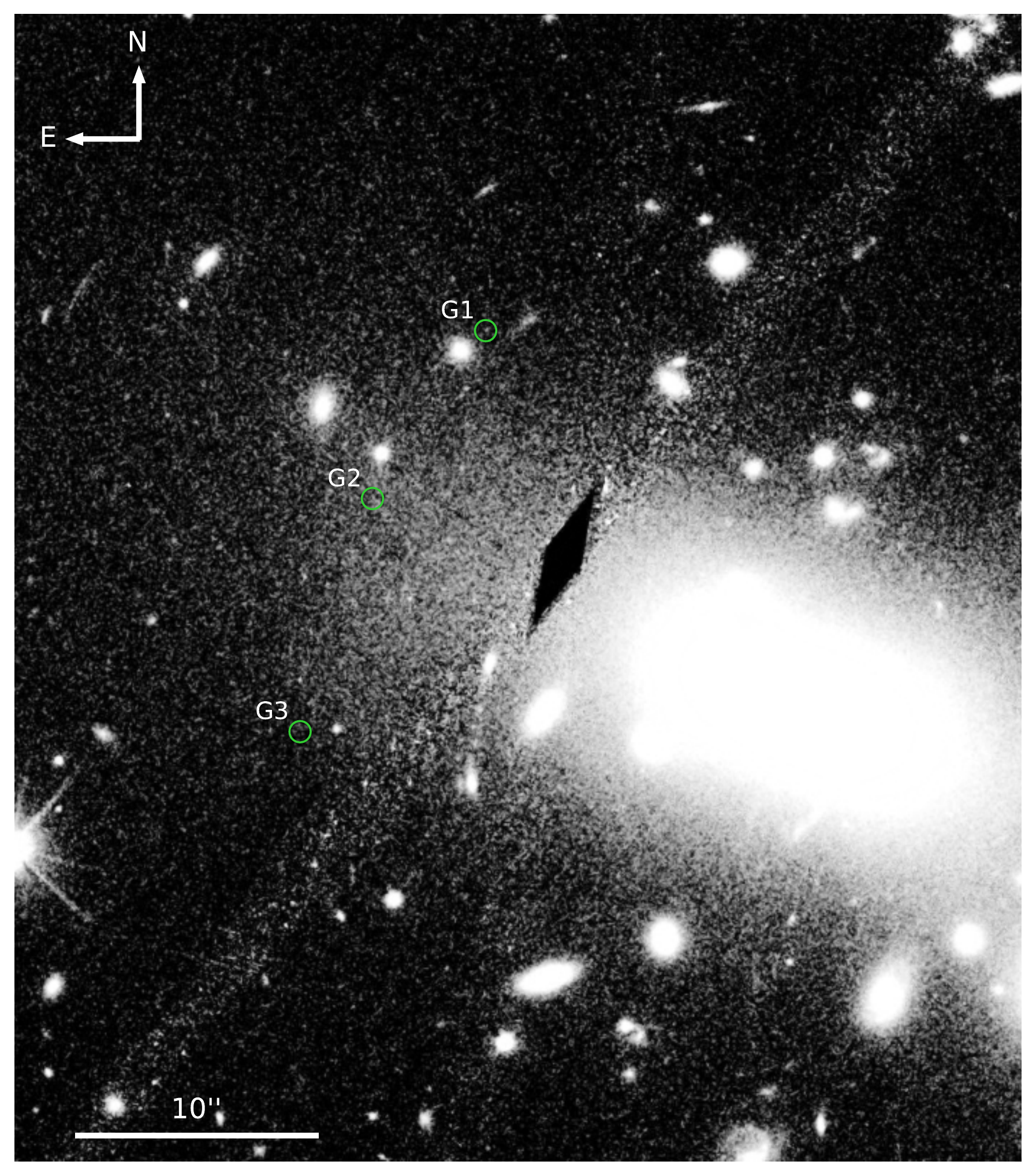}
\caption{\textbf{Blue image of part of RX\,J2129. }The blue frame from the color-composite image of part of RX\,J2129 (Figure \ref{fig:imglarge}), with three images of the $z=9.51$ galaxy circled in green. This image is a composite of the HST ACS F606W + F814W images. The broad bands are diffraction spikes from foreground stars. The black diamond is an artefact caused by a chip gap in the HST ACS camera. }
\label{fig:rxj2129_blue}
\end{figure*}


\end{document}